\RequirePackage[mathlines]{lineno}
\documentclass[amsmath,amssymb,twocolumn,prd,floatfix,showpacs, nofootinbib]{revtex4-1}
\usepackage{tabularx}  
\usepackage{bm, color} 
\usepackage{overpic,subfigure} 
\usepackage{multirow}
\usepackage{array}
\usepackage{dcolumn} 
\usepackage[symbol]{footmisc}
\usepackage{epstopdf}
\usepackage{lineno}
\usepackage{hyperref}
\usepackage{amsmath}
\usepackage{float}

\newcommand{\BESIIIorcid}[1]{\href{https://orcid.org/#1}{\hspace*{0.1em}\raisebox{-0.45ex}{\includegraphics[width=1em]{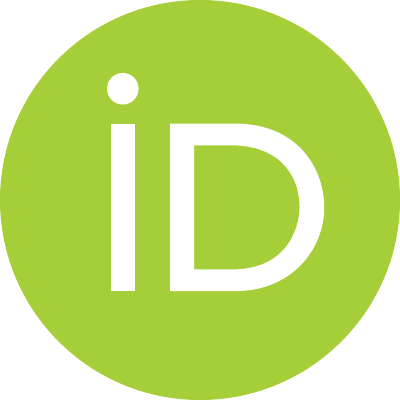}}}}

\hypersetup{hypertex=true,
	        colorlinks = true,
            linkcolor = blue,
            urlcolor = blue,
            citecolor = blue}

\begin{document}

\title{\boldmath Measurement of the branching fractions for the decays $\chi_{cJ}\to \phi \omega \eta$ and $\chi_{cJ}\to \phi \omega \eta^{\prime}$ }

\author{
M.~Ablikim$^{1}$\BESIIIorcid{0000-0002-3935-619X},
M.~N.~Achasov$^{4,c}$\BESIIIorcid{0000-0002-9400-8622},
P.~Adlarson$^{83}$\BESIIIorcid{0000-0001-6280-3851},
X.~C.~Ai$^{89}$\BESIIIorcid{0000-0003-3856-2415},
C.~S.~Akondi$^{31A,31B}$\BESIIIorcid{0000-0001-6303-5217},
R.~Aliberti$^{39}$\BESIIIorcid{0000-0003-3500-4012},
A.~Amoroso$^{82A,82C}$\BESIIIorcid{0000-0002-3095-8610},
Q.~An$^{78,65,\dagger}$,
Y.~H.~An$^{89}$\BESIIIorcid{0009-0008-3419-0849},
M.~S.~Anderson$^{39}$\BESIIIorcid{0009-0008-1550-2632}
% ...
\\
\vspace{0.2cm}
(BESIII Collaboration)\\
\vspace{0.2cm} {\it
$^{1}$ Institute of High Energy Physics, Beijing 100049, People's Republic of China\\
% ...
$^{4}$ Budker Institute of Nuclear Physics SB RAS (BINP), Novosibirsk 630090, Russia\\
% ...
$^{31}$ INFN Sezione di Ferrara, (A)INFN Sezione di Ferrara, I-44122, Ferrara, Italy; (B)University of Ferrara, I-44122, Ferrara, Italy\\
% ...
$^{39}$ Johannes Gutenberg University of Mainz, Johann-Joachim-Becher-Weg 45, D-55099 Mainz, Germany\\
% ...
$^{65}$ State Key Laboratory of Particle Detection and Electronics, Beijing 100049, Hefei 230026, People's Republic of China\\
% ...
$^{78}$ University of Science and Technology of China, Hefei 230026, People's Republic of China\\
% ...
$^{82}$ University of Turin and INFN, (A)University of Turin, I-10125, Turin, Italy; (B)University of Eastern Piedmont, I-15121, Alessandria, Italy; (C)INFN, I-10125, Turin, Italy\\
$^{83}$ Uppsala University, Box 516, SE-75120 Uppsala, Sweden\\
% ...
$^{89}$ Zhengzhou University, Zhengzhou 450001, People's Republic of China
\\
\vspace{0.2cm}
$^{\dagger}$ Deceased\\
% ...
$^{c}$ Also at the Novosibirsk State University, Novosibirsk, 630090, Russia\\
% ...
}

}		

\begin{abstract}
  Using a sample of $(2712.4 \pm 14.3) \times 10^{6}$ $\psi(3686)$ events collected with the BESIII detector at the BEPCII collider, we report the first observation and measurement of the branching fractions for the decays $\chi_{cJ}\to \phi \omega \eta$ and $\chi_{cJ} \to \phi \omega \eta^{\prime}$ ($J = 0,1,2$) with statistical significance greater than $5\sigma$. The branching fractions are determined to be $\mathcal B(\chi_{c0} \to \phi \omega \eta) = (5.43 \pm 0.23 \pm 0.52) \times 10^{-4}$, $\mathcal B(\chi_{c1} \to \phi \omega \eta) = (2.51 \pm 0.14 \pm 0.24) \times 10^{-4}$ and $\mathcal B(\chi_{c2} \to \phi \omega \eta) = (3.03 \pm 0.15 \pm 0.28) \times 10^{-4}$  for the $\eta$ mode, and
  $\mathcal B(\chi_{c0} \to \phi \omega \eta^{\prime}) = (1.23 \pm 0.23 \pm 0.15) \times 10^{-4}$, $\mathcal B(\chi_{c1} \to \phi \omega \eta^{\prime}) = (1.09 \pm 0.17 \pm 0.13) \times 10^{-4}$ and $\mathcal B(\chi_{c2} \to \phi \omega \eta^{\prime}) = (0.98 \pm 0.17 \pm 0.11) \times 10^{-4}$  for the $\eta^{\prime}$ mode, where the first uncertainties are statistical and the second systematic.
\end{abstract}

\maketitle

\section{Introduction}

Experimental studies of charmonium states and their decay properties are important to test quantum chromodynamics models~\cite{ref::work1, ref::work2, ref::work4}. The $\chi_{cJ}~(J = 0,1,2)$ mesons are the $^{3}P_{J}$ charmonium states, and have not been studied as extensively as the vector charmonium states.
Parity conservation forbids the direct production of $\chi_{cJ}$ states in $e^+e^-$ annihilation via a single virtual photon. Production through two virtual photons is allowed but strongly suppressed, resulting in low production rates. In contrast, radiative decays of $\psi(3686)$ into $\chi_{cJ}$ have branching fractions (BFs) of about 9\%~\cite{ref::pdg2025} and provide large and clean $\chi_{cJ}$ samples. Currently, the sum of all measured BFs of $\chi_{cJ}$ decays is still far below unity, and knowledge of exclusive decays remains limited. According to the current world-average measurements, nearly 30\% of the $\chi_{cJ}$ decay BFs remain unknown. Further searches for unknown $\chi_{cJ}$ decays will be helpful for a better understanding of their decay properties.

To date, only a few measurements have been reported for $\chi_{cJ} \to VVP$ decays, where $V$ and $P$ denote vector and pseudoscalar mesons~\cite{BESIII:2019ngm}, respectively. In particular, the channels $\chi_{cJ} \to \phi \omega \eta$ and $\chi_{cJ} \to \phi \omega \eta^{\prime}$ have not yet been studied experimentally. The BESIII Collaboration has collected $(2712.4 \pm 14.3)\times10^{6}$ $\psi(3686)$ events~\cite{ref::work3} since 2008, providing an excellent opportunity to study previously unexplored $\chi_{cJ}$ decay modes with the large $\chi_{cJ}$ samples produced in radiative $\psi(3686)$ decays. In this paper, we present the first measurements of the branching fractions of $\chi_{cJ}\to \phi \omega \eta$ and $\chi_{cJ}\to \phi \omega \eta^{\prime}$ using this data sample.


\section{BESIII DETECTOR AND MONTE CARLO SIMULATION}
\label{sec:BES}
The BESIII detector~\cite{ref::detector} records symmetric $e^+e^-$ collisions provided by the BEPCII storage ring~\cite{ref::collider}
in the center-of-mass energy range from 1.84 to 4.95~GeV, with a peak luminosity of $1.1 \times 10^{33}\;\text{cm}^{-2}\text{s}^{-1}$ achieved at $\sqrt{s} = 3.773\;\text{GeV}$.

The cylindrical core of the BESIII detector covers 93\% of the full solid angle and consists of a helium-based multilayer drift chamber~(MDC),
a plastic scintillator time-of-flight system~(TOF), and a CsI (Tl) electromagnetic calorimeter~(EMC), all enclosed in a superconducting
solenoidal magnet providing a 1.0~T magnetic field. The solenoid is supported by an octagonal flux-return yoke with resistive plate counter muon
identification modules interleaved with steel. The charged-particle momentum resolution at $1~{\rm GeV}/c$ is $0.5\%$, and the ${\rm d}E/{\rm d}x$
resolution is $6\%$ for electrons from Bhabha scattering. The EMC measures photon energies with a resolution of $2.5\%$ ($5\%$) at $1$~GeV in the barrel
(end cap) region. The time resolution in the TOF barrel region is 68~ps, while that in the end cap region is 110~ps.
The end cap TOF system was upgraded in 2015 using multi-gap resistive plate chamber technology, providing a time resolution of 60~ps~\cite{Tof1,Tof2,Tof3}.
About $83\%$ of the data used  in this analysis benefit from this upgrade.

Simulated data samples are produced with a {\sc
geant4}-based~\cite{Geant4} Monte Carlo (MC) package, which
includes the geometric description~\cite{Geometric} of the BESIII detector and the
detector response. These samples are used to determine detection efficiencies
and to estimate backgrounds. The simulation models the beam
energy spread and initial state radiation (ISR) in $e^+e^-$
annihilations with the generator {\sc
kkmc}~\cite{Jadach01, Jadach02}.
The inclusive MC sample includes the production of the
$\psi(3686)$ resonance, the ISR production of the $J/\psi$, and
the continuum processes incorporated in {\sc
kkmc}~\cite{Jadach01}.
All particle decays are modeled with {\sc
evtgen}~\cite{Lange01,Lange02} using branching fractions
either taken from the
Particle Data Group (PDG)~\cite{ref::pdg2025}, when available,
or otherwise estimated with {\sc lundcharm}~\cite{Lundcharm00, Lundcharm01}.
Final state radiation
from charged final state particles is incorporated using the {\sc photos} package~\cite{PHOTOS}.
An inclusive MC sample containing $2.7\times10^{9}$ generic $\psi(3686)$ events is used to study backgrounds.
To determine the detection efficiency for signal events, 500,000 signal MC samples are generated with a modified data-driven generator BODY3~\cite{Lange01,BODY3},
which can simulate contributions from different intermediate states in data for a given three-body final state, as discussed in Sec.~\ref{bfs}.
The $\psi(3686) \to \gamma \chi_{cJ} ~(J = 0,1,2)$ decays are generated with the P2GCJ~\cite{Lange01} models. 
The decay $\phi \to K^{+}K^{-}$ is generated by the VSS~\cite{VSS} model (vector to a pair of pseudoscalar mesons).
The decay $\omega \to \pi^{+}\pi^{-}\pi^{0}$ is generated with the model in ref~\cite{Lange01}.
The decay $\eta^{\prime} \to \pi^{+}\pi^{-}\eta$ is generated with the Etap2pipieta~\cite{ETAP2PIPIETA} model.
The other signal decays are generated according to phase space (PHSP).

\section{EVENT SELECTION}
\label{sec:selection}

The $\phi$, $\omega$, $\eta^{\prime}$, $\eta$, and $\pi^{0}$ mesons are reconstructed via their dominant decay modes $\phi \to K^+ K^-$, $\omega \to \pi^+ \pi^- \pi^0$, $\eta^{\prime} \to \pi^+ \pi^- \eta$, $\eta \to \gamma \gamma$ and $\pi^0 \to \gamma \gamma$, respectively. 
For the $\eta^{\prime} \to \pi^+ \pi^- \gamma$ decay, the backgrounds are difficult to remove completely, resulting in a much lower signal significance than that of the $\pi^+ \pi^- \eta$ mode. Therefore, only $\pi^+ \pi^- \eta$ is used for $\eta^{\prime}$ reconstruction.
The BFs of $\chi_{cJ} \to \phi \omega \eta$ and $\chi_{cJ} \to \phi \omega \eta^{\prime}$ are measured via the radiative decays $\psi(3686) \to \gamma \chi_{cJ}$. The signal events are required to contain four and six charged tracks with zero net charge for the $\phi\omega\eta$ and $\phi\omega\eta^{\prime}$ modes, respectively, and at least five photon candidates.

Photon candidates are identified using isolated showers in the EMC. The deposited energy of each shower must be more than 25~MeV in the barrel region ($|\cos \theta|< 0.80$), where $\theta$ denotes the polar angle defined with respect to the MDC symmetry axis, and more than 50~MeV in the end cap region ($0.86 <|\cos \theta|< 0.92$). To exclude showers that originate from charged tracks, the opening angle between the EMC shower and the position of the closest charged track at the EMC must be greater than 10 degrees as measured from the interaction point (IP). To suppress electronic noise and showers unrelated to the event, the difference between the EMC time and the event start time is required to be within [0, 700]\,ns.

Charged tracks in an event are required to have a polar angle $\theta$ within the MDC acceptance $|\cos \theta|< 0.93$, and a distance of closest approach to the IP within 10 cm along the beam direction and 1 cm in the plane transverse to the beam direction. Particle identification~(PID) for charged tracks combines measurements of d$E$/d$x$ in the MDC and flight time in the TOF to form likelihoods $\mathcal{L}(h)$ for each hadron hypothesis $h$ ~($h$ = $\pi$, $K$). Charged tracks with $\mathcal{L}(K)>\mathcal{L}(\pi)$ are assigned as kaons, while those with $\mathcal{L}(\pi)>\mathcal{L}(K)$ are assigned as pions.

The $\phi$ candidates are selected by requiring the $K^+ K^-$ invariant mass to satisfy $M(K^{+}K^{-}) \in [1.005,\,1.035]\ \text{GeV}/c^{2}$. To estimate the background contribution, we additionally analyze the events selected from the one-dimensional (1D) $\phi$ meson sideband region, defined as the invariant mass interval $M(K^{+}K^{-}) \in [1.045,\,1.075]\ \text{GeV}/c^{2}$.
To suppress background and improve the mass resolution, a six-constraint (6C) kinematic fit is performed under the decay hypothesis $\psi(3686) \to \gamma \chi_{cJ}$, $\chi_{cJ} \to {\rm signal}$. The fit constrains the total four-momentum of the final-state particles to that of the colliding beams and the invariant masses of the two photon pairs to the nominal $\pi^{0}$ and $\eta$ masses, respectively~\cite{ref::pdg2025}.
In each event, if more than one combination survives, the one with the smallest $\chi_{\rm 6C}^{2}$ of the 6C kinematic fit is retained. The requirement 
$\chi^2_{\rm 6C}<45$ is determined by maximizing the figure of merit (FOM), defined as ${\rm FOM} = S/\sqrt{S+B}$, where $S$ is the number of events from signal MC sample normalized to the preliminary signal yield obtained from data before applying the $\chi^2_{\rm 6C}$ requirement, $B$ is the number of background events from the inclusive MC sample, and has been normalized to the size of the data.
To suppress the background with four or six photons in the final state, a 4C kinematic fit is performed by imposing the energy-momentum conservation under the hypotheses with different numbers of photons. The combination with the smallest $\chi_{\rm 4C}^{2}$ is retained for each channel. In addition, the following requirements are imposed:
\begin{gather*}
\chi^{2}_{\rm 4C}(K^+ K^- \pi^+ \pi^- 5\gamma) ~\textless~ \chi^{2}_{\rm 4C}(K^+ K^- \pi^+ \pi^- 4\gamma),\\
\chi^{2}_{\rm 4C}(K^+ K^- \pi^+ \pi^- 5\gamma) ~\textless~ \chi^{2}_{\rm 4C}(K^+ K^- \pi^+ \pi^- 6\gamma),\\
\chi^{2}_{\rm 4C}(K^+ K^- \pi^+ \pi^+ \pi^- \pi^- 5\gamma) ~\textless~ \chi^{2}_{\rm 4C}(K^+ K^- \pi^+ \pi^+ \pi^- \pi^- 4\gamma),\\
\chi^{2}_{\rm 4C}(K^+ K^- \pi^+ \pi^+ \pi^- \pi^- 5\gamma) ~\textless~ \chi^{2}_{\rm 4C}(K^+ K^- \pi^+ \pi^+ \pi^- \pi^- 6\gamma).
\end{gather*}

The potential background components from $\psi(3686)$ decays are studied by analyzing the inclusive MC sample with the generic event type analysis tool TopoAna~\cite{topo}. For the $\chi_{cJ} \to \phi \omega \eta$ channel, to suppress the background from the decay $\psi(3686) \to \pi^{+}\pi^{-}J/\psi$, the recoil mass of $\pi^{+}\pi^{-}$, $RM(\pi^{+}\pi^{-})$, is required to satisfy $|RM(\pi^{+}\pi^{-}) - M(J/\psi)| \textgreater 0.012$ GeV/$c^{2}$, where $M(J/\psi)$ is the nominal mass of $J/\psi$~\cite{ref::pdg2025}. To suppress the background from the decay $\psi(3686) \to \pi^{0}\pi^{0}J/\psi$, we veto events in the region where $M(K^{+}K^{-}\pi^{+}\pi^{-}\eta)$ lies between $3.08$ and $3.15$ GeV$/c^{2}$, and where the recoil mass of 4$\gamma$, $RM(4\gamma)$, closest to the $J/\psi$ mass falls between $3.08$ GeV/$c^{2}$ and $3.12$ GeV/$c^{2}$.
For $\chi_{cJ} \to \phi \omega \eta$ and $\chi_{cJ} \to \phi \omega \eta^{\prime}$, no peaking background 
is found in the $\chi_{cJ}$ signal region while simultaneously containing a $\omega$ candidate.
In addition, the possible continuum background contribution is evaluated by examining a data sample taken at $\sqrt s =$ 3.650~GeV, corresponding to an integrated luminosity of 445.49~pb$^{-1}$~\cite{ref::work3}. Since no events survive the selection criteria within the $\chi_{cJ}$ signal regions, the continuum contribution is considered negligible.

\section{Branching fraction}
\label{bfs}
\subsection{The decay $\chi_{cJ} \to \phi \omega \eta$}
\label{4kfit}
\begin{figure*}[htpt]
	\centering
	\includegraphics[width=14.2cm]{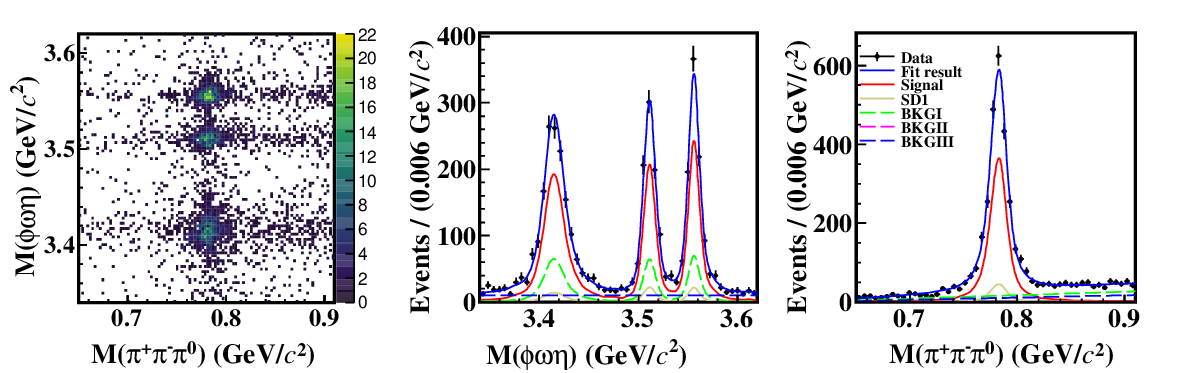}
	\includegraphics[width=14.2cm]{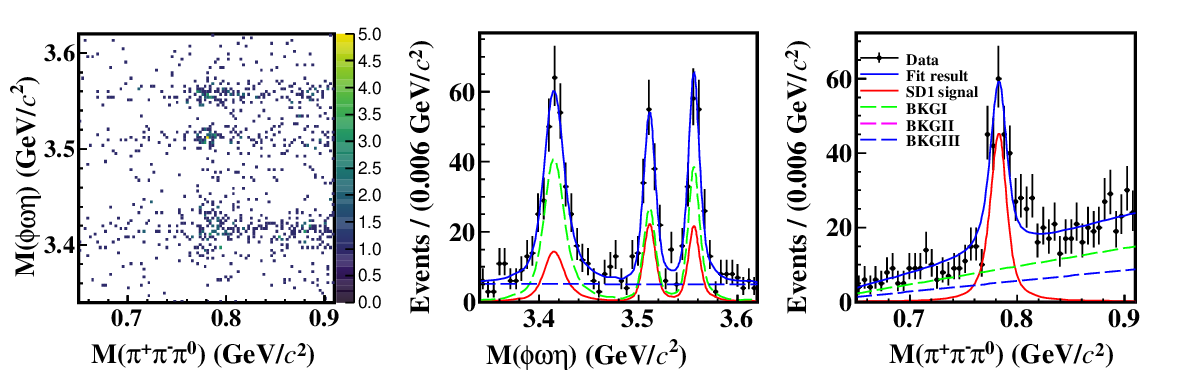}
	\caption{Distribution of $M(\phi \omega \eta)$ versus $M(\pi^+\pi^-\pi^0)$ (left column), and projections of the 2D simultaneous fit to the $M(\phi \omega \eta)$ (middle column) and $M(\pi^+\pi^-\pi^0)$ (right column) distributions in the $\phi$ signal (top row) and sideband region (bottom row). Data are shown as the dots with error bars. For both the top and bottom figures, the blue solid curves are the fit results, the red solid curves are the fitted signal shapes, the green dashed curves are the BKGI events, the pink dashed curves are the BKGII events, and the blue dashed curves are the BKGIII events. For the top figures, the brown solid curves are the normalized sideband contributions.
}
	\label{fit:phi_sig}
\end{figure*}

To determine the signal yields and constrain the nonresonant background in $\chi_{cJ} \to \phi \omega \eta$ events, a simultaneous two-dimensional (2D) unbinned maximum likelihood  fit is performed to the $M(\phi \omega \eta)$ versus $M(\pi^{+}\pi^{-}\pi^{0})$ distributions of selected data candidates in the $\phi$ signal and sideband regions. The $\phi$ signal region is defined as $M(K^{+}K^{-}) \in [1.005,\,1.035]\ \text{GeV}/c^{2}$ and the sideband region is defined as $M(K^{+}K^{-}) \in [1.045,\,1.075]\ \text{GeV}/c^{2}$.
In the 2D simultaneous fit, 
the selected candidates are categorized into four components:
``Signal", in which the $\phi\omega\eta$ and $\pi^+\pi^-\pi^0$ candidates originate from $\chi_{cJ}$ and $\omega$ decays, respectively;
``BKGI'', in which the $\phi\omega\eta$ candidates originate from $\chi_{cJ}$ decays, while the $\pi^+\pi^-\pi^0$ candidates arise from combinatorial backgrounds;
``BKGII'', in which the $\phi\omega\eta$ candidates arise from combinatorial backgrounds, while the $\pi^+\pi^-\pi^0$ candidates originate from $\omega$ decays;
``BKGIII'', in which both $\phi\omega\eta$ and $\pi^+\pi^-\pi^0$ candidates arise from combinatorial backgrounds, including misreconstructed events with different final states.
Therefore, the probability density functions (PDFs) of Signal, BKGI, BKGII, and BKGIII are constructed as follows:
\begin{itemize}
\item Signal: $S(x) \times S(y)$,
\item BKGI: $S(x) \times C(y)$,
\item BKGII: $C(x) \times S(y)$,
\item BKGIII: $C(x) \times C(y)$.
\end{itemize}
Here, $x$ and $y$ represent $M(\phi \omega \eta)$ and $M(\pi^{+}\pi^{-}\pi^{0})$, respectively, corresponding to the two dimensions of the 2D fit. $S(i)$ denotes the signal shapes derived from the signal MC samples convolved with a Gaussian function. The Gaussian parameters in the $\phi$ signal-region fit are shared with those in the sideband-region fit. $C(i)$ denotes a second-order Chebyshev polynomial, where $i$ labels the fit dimensions.
Events in the $\phi$ sideband region are normalized using a scale factor $f$, which is determined to be 1.0 by integrating the background function over the $\phi$ signal and sideband regions.
Figure~\ref{fit:phi_sig} shows the projections of the 2D simultaneous fit results to the signal and sideband candidate events.
The signal yields are summarized in Table~\ref{tab:branching_fraction}.
The statistical significance of each $\chi_{c0,1,2}\to\phi\omega\eta$ signal exceeds $5\sigma$. The significance is evaluated from the change in the log-likelihood value between fits with and without the signal component, taking into account the corresponding change in the number of free parameters ($\Delta ndf=1$).

\begin{figure*}[htpt]
\centering
		\includegraphics[width=0.255\textwidth]{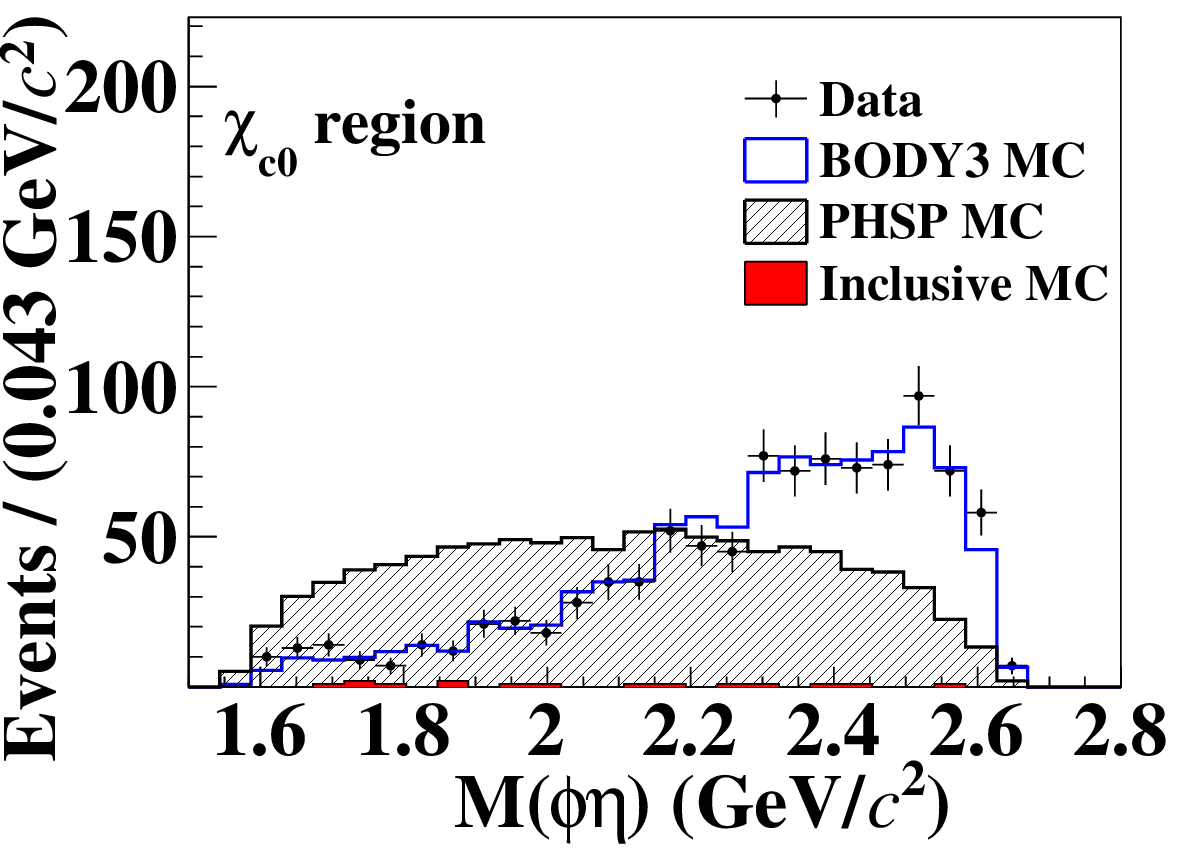}
		\includegraphics[width=0.255\textwidth]{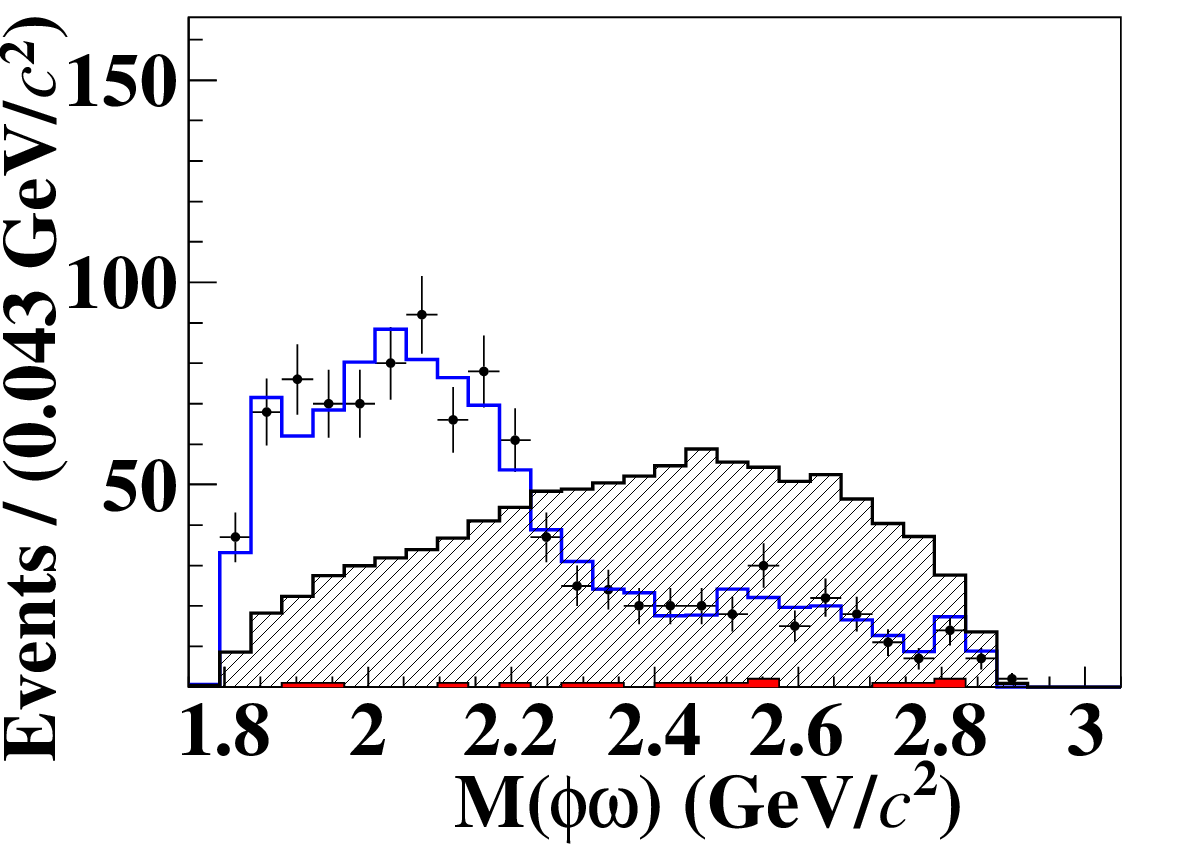}
		\includegraphics[width=0.255\textwidth]{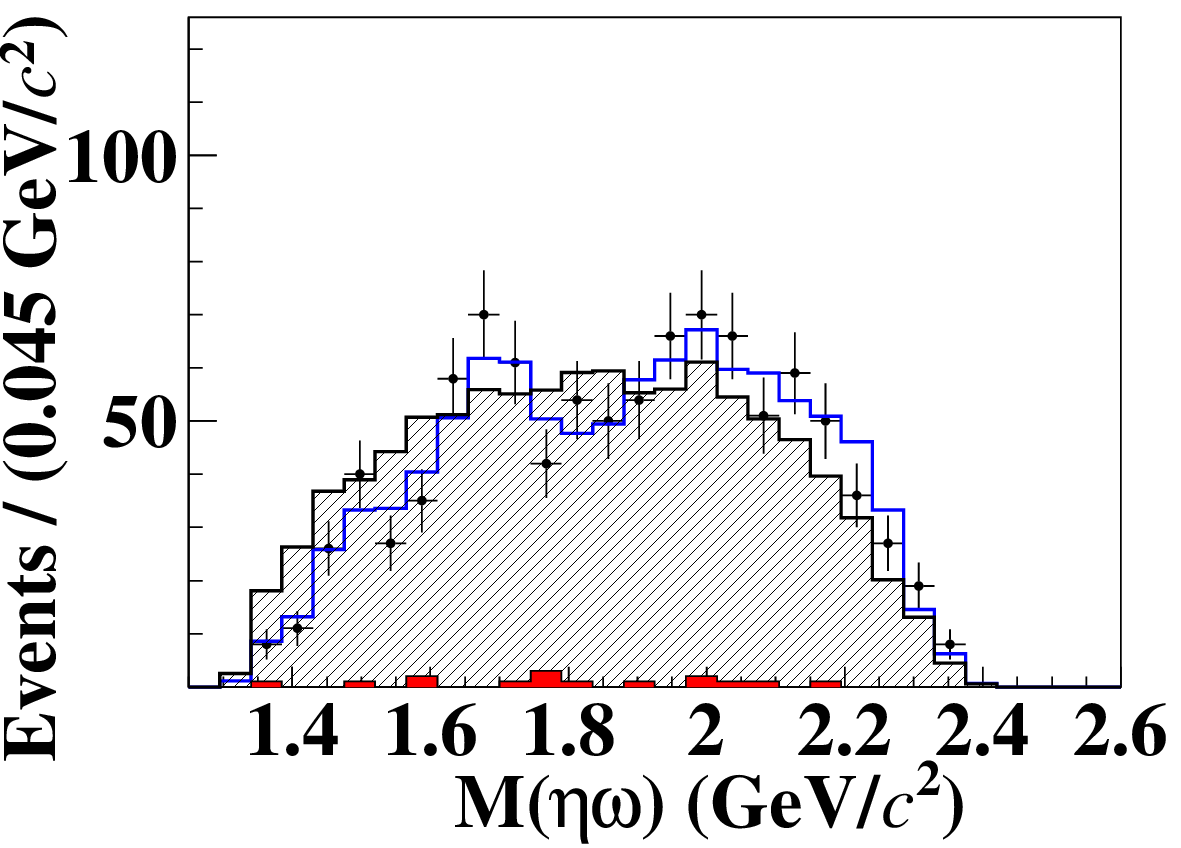}\\
		\includegraphics[width=0.255\textwidth]{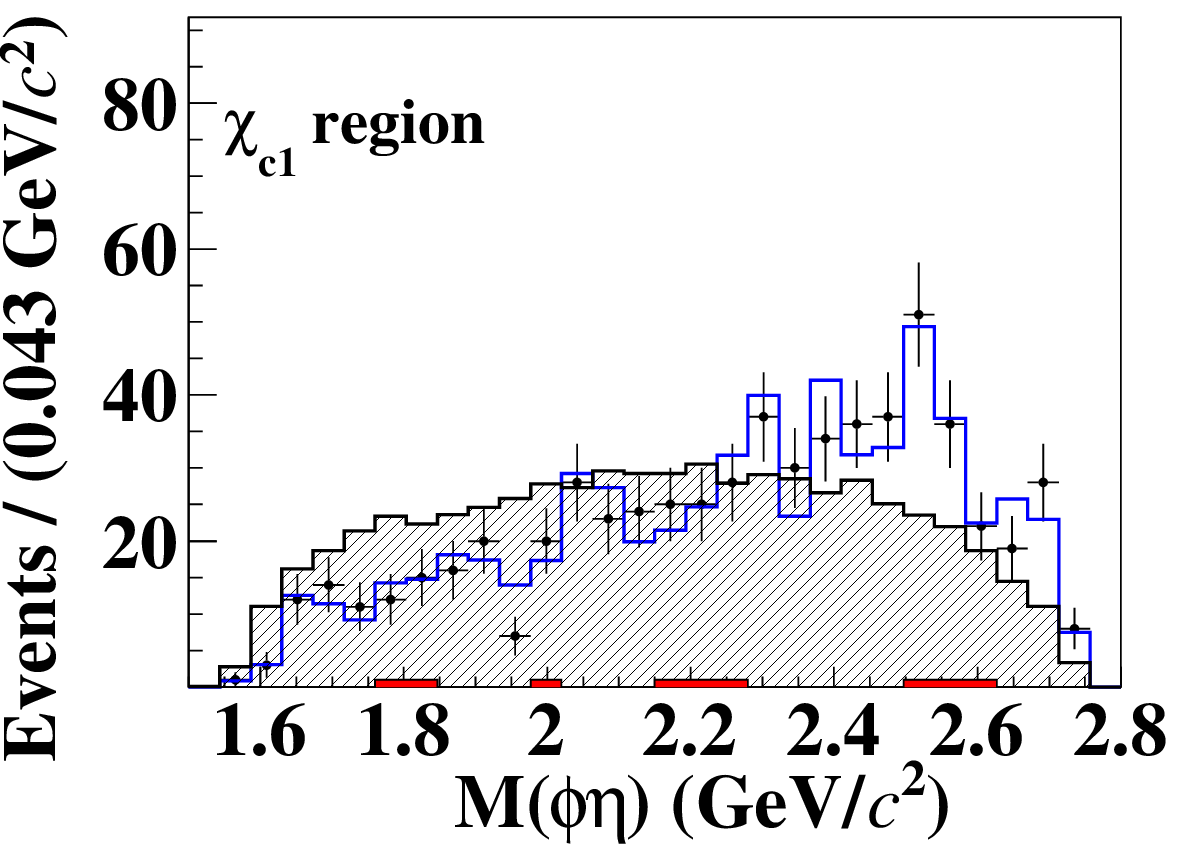}
		\includegraphics[width=0.255\textwidth]{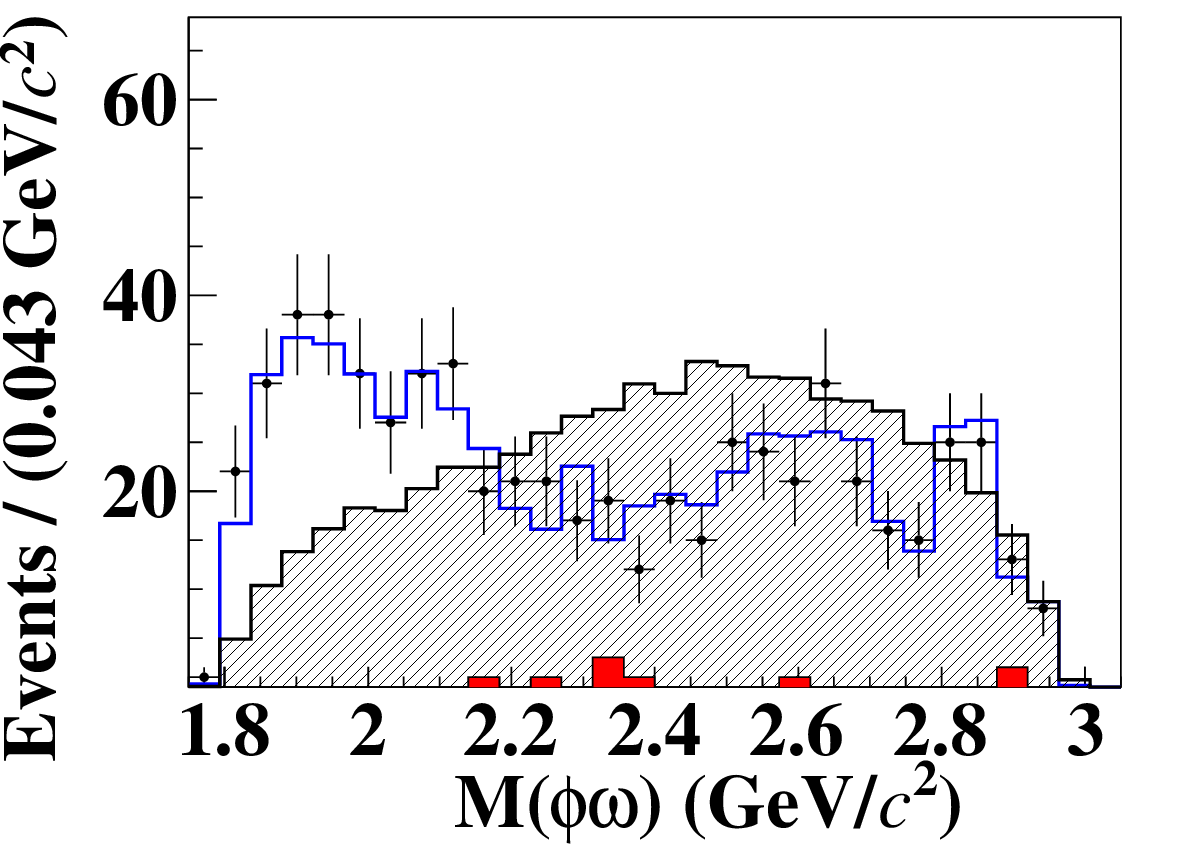}
		\includegraphics[width=0.255\textwidth]{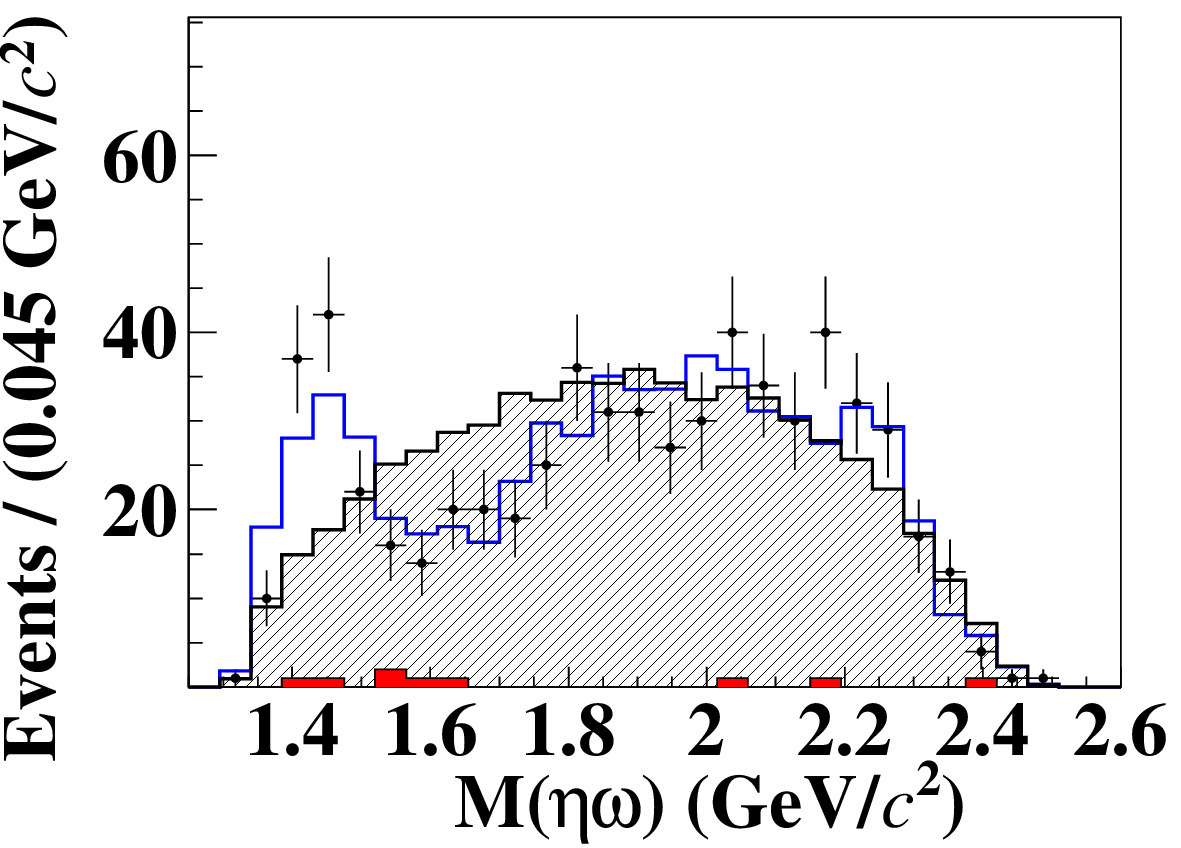}\\
		\includegraphics[width=0.255\textwidth]{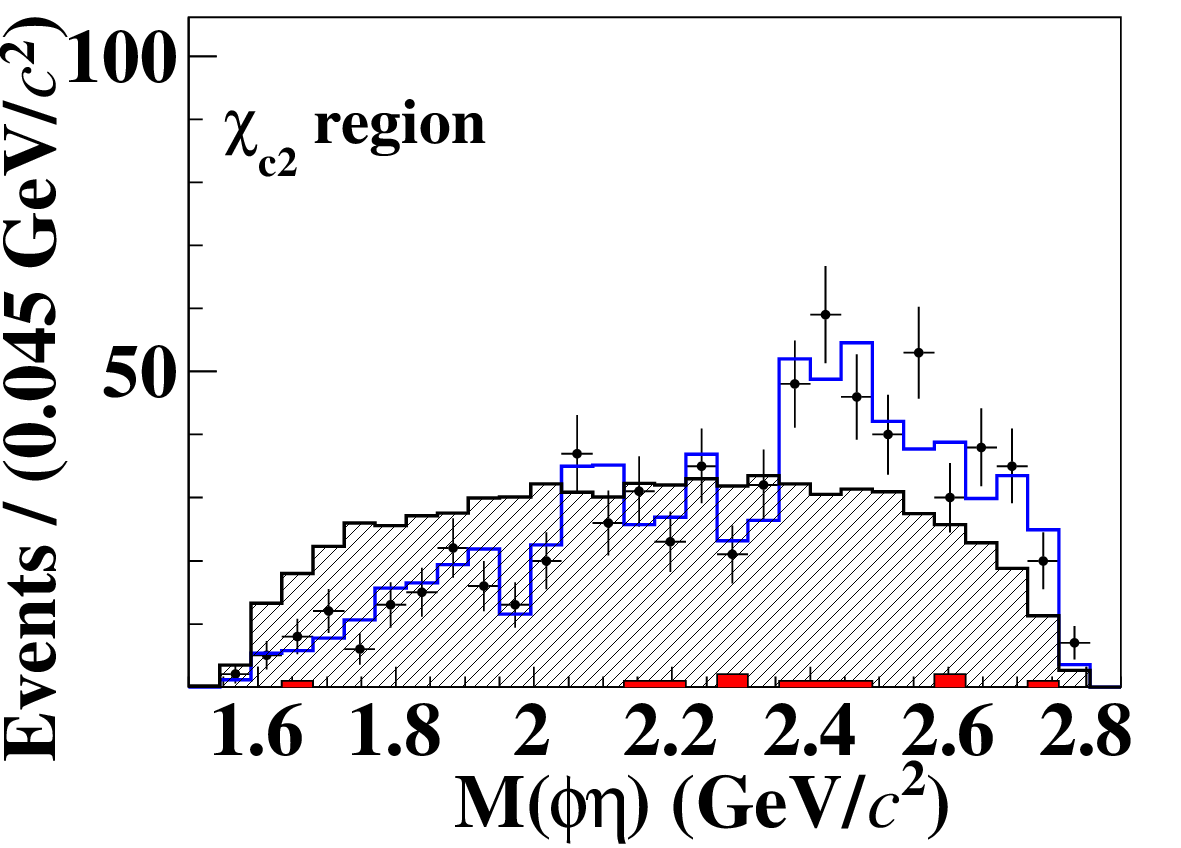}
		\includegraphics[width=0.255\textwidth]{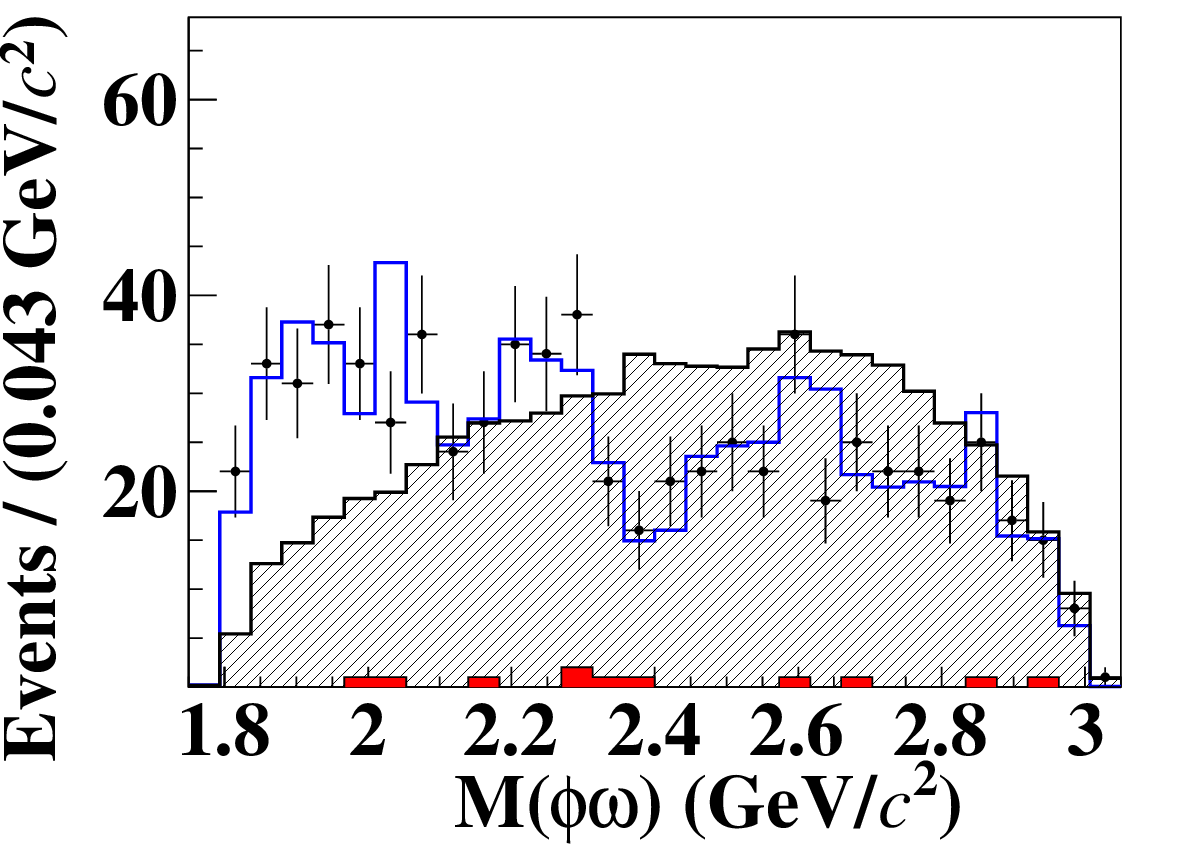}
		\includegraphics[width=0.255\textwidth]{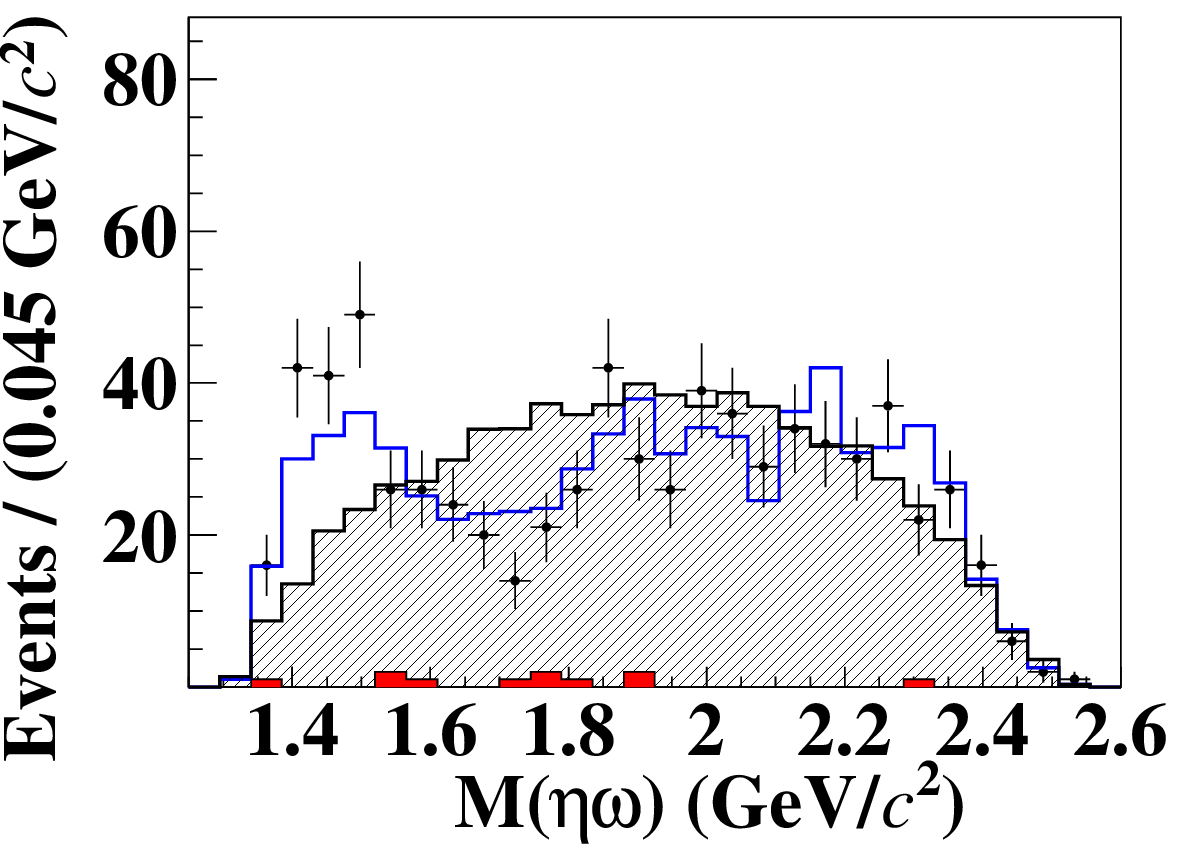}\\
		\caption{Invariant mass distributions for all possible two-body combinations of the accepted $\chi_{cJ} \to \phi \omega \eta$ candidates. The top, middle, and bottom rows are for events in the $\chi_{c0}$ mass region [3.385, 3.445] GeV/$c^{2}$, the $\chi_{c1}$ mass region [3.490, 3.530] GeV/$c^{2}$, and the $\chi_{c2}$ mass region [3.536, 3.576] GeV/$c^{2}$, respectively. The dots with error bars are data, blue curves are the BODY3 MC samples, red histograms are the background contributions as estimated by the inclusive MC samples, and the black histograms are the PHSP MC samples. The PHSP MC and BODY3 MC are normalized to the signal yield in data. Events have been required to be within the $\omega$ signal region defined as $|M(\pi^{+}\pi^{-}\pi^{0}) - M(\omega)|~ \textless~ 0.024$ GeV/$c^{2}$.
			\label{comparision_phi}
}
\end{figure*}

The existence of potential intermediate states in the $\phi \omega \eta$ final state is investigated by scrutinizing the invariant mass distributions. Figure~\ref{comparision_phi} shows the invariant mass distributions of the different two-body combinations for the $\chi_{c0,1,2}$ signal regions, after subtracting the background contributions using the inclusive MC samples. Clear structures are observed in the $\omega \eta$ system close to the $\omega \eta$ threshold in $\chi_{c1,2}$ signal regions, 
which are not well described by the PHSP model.
Therefore, 
the modified data-driven generator BODY3~\cite{Lange01,BODY3} 
is employed 
to generate signal MC samples to avoid bias in the efficiency estimation.

\begin{figure*}[htpt]
\centering
\includegraphics[width=0.28\textwidth]{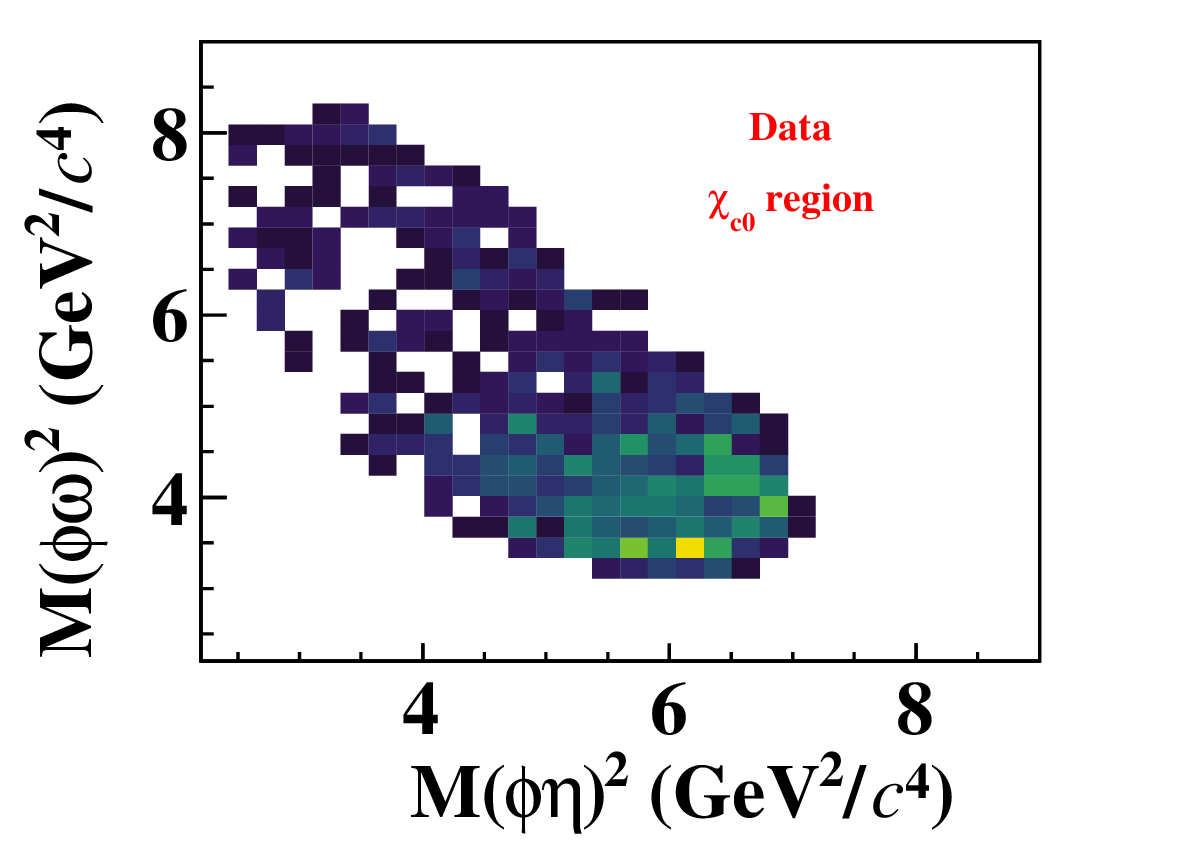}
\includegraphics[width=0.28\textwidth]{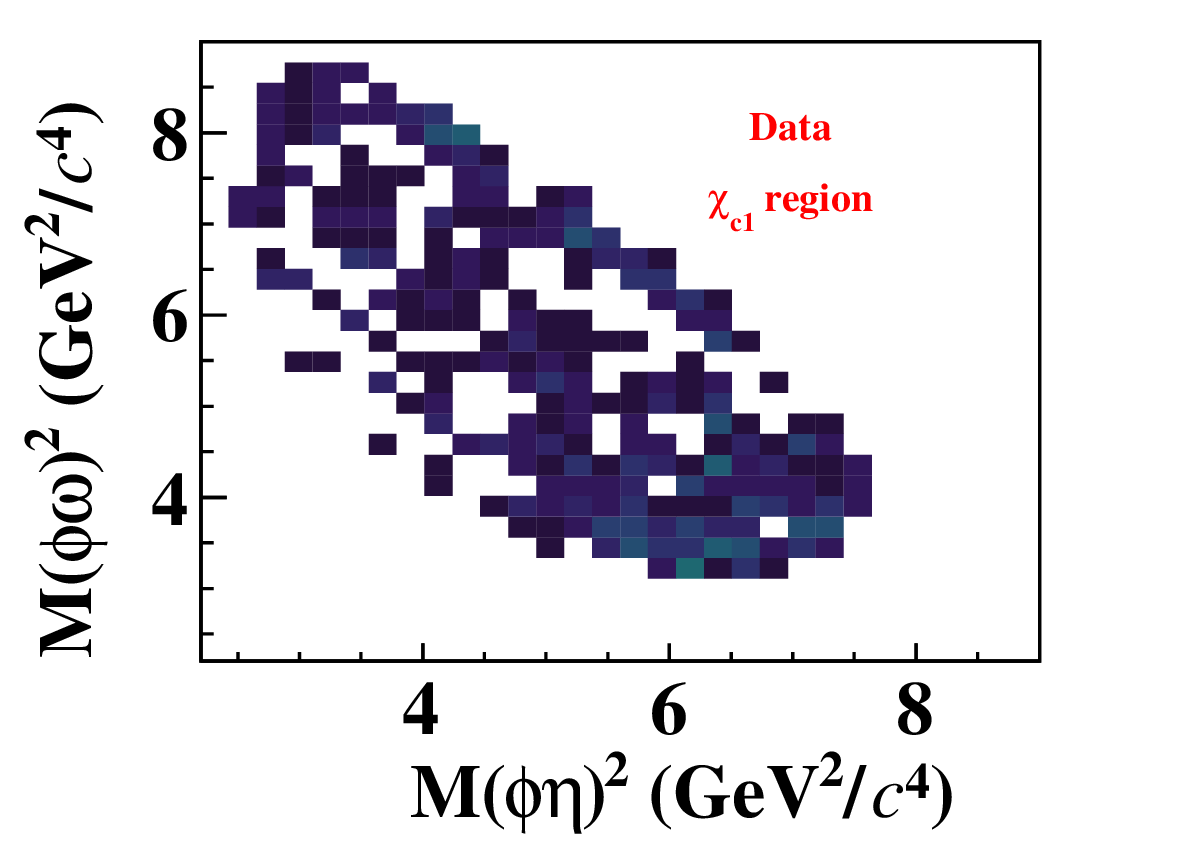}
\includegraphics[width=0.28\textwidth]{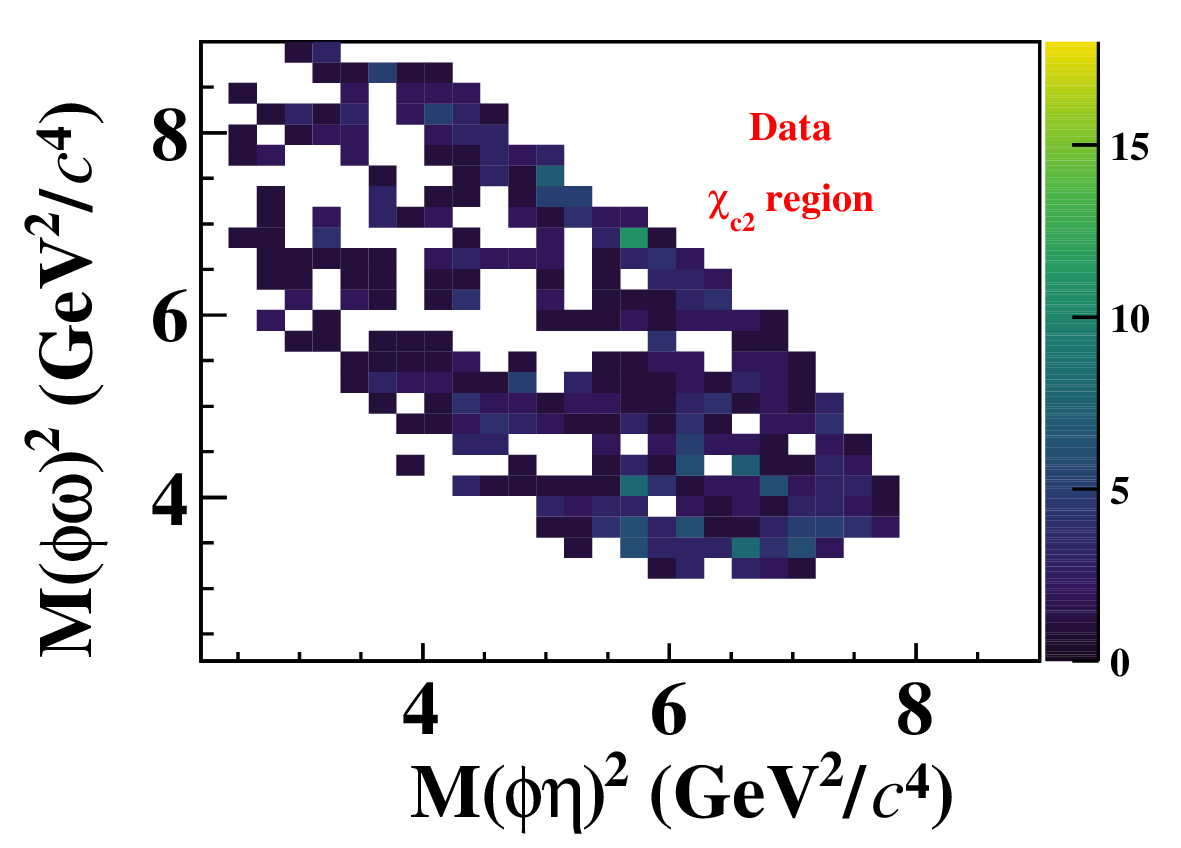}\\
\includegraphics[width=0.28\textwidth]{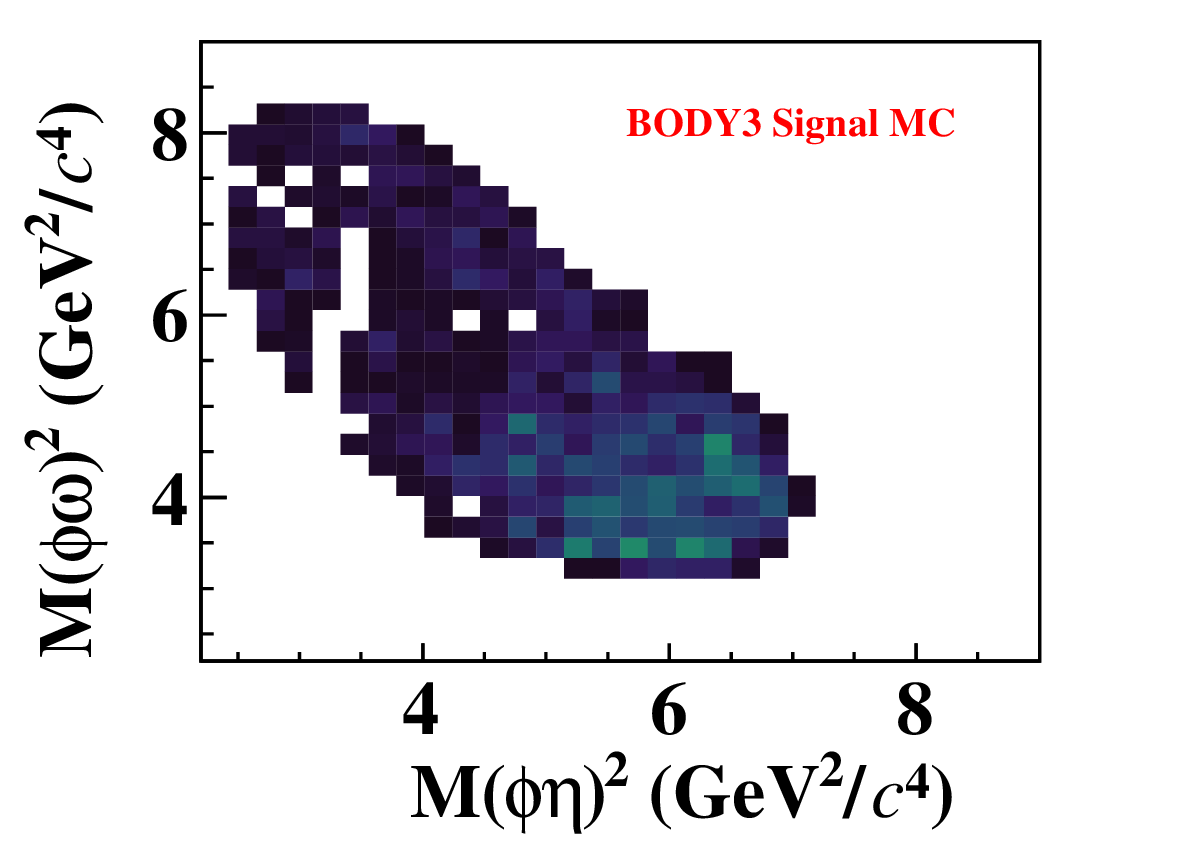}
\includegraphics[width=0.28\textwidth]{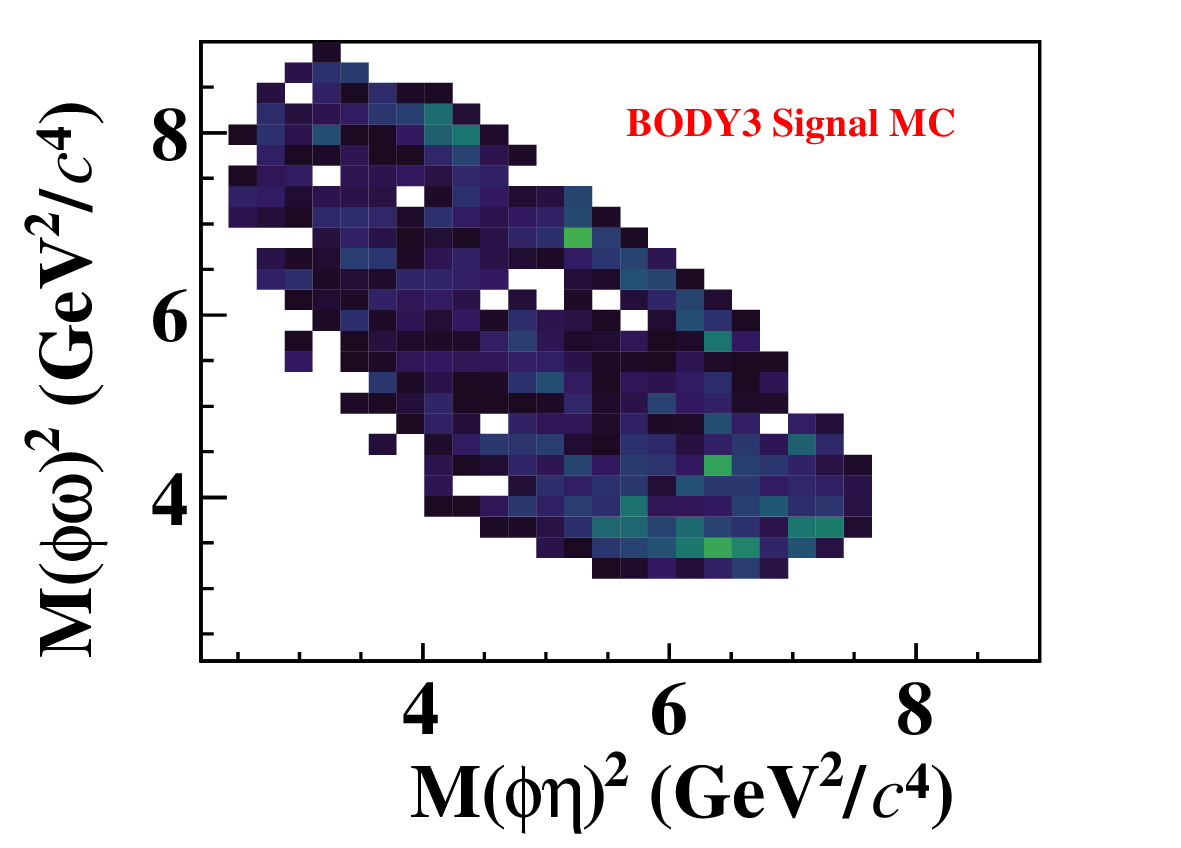}
\includegraphics[width=0.28\textwidth]{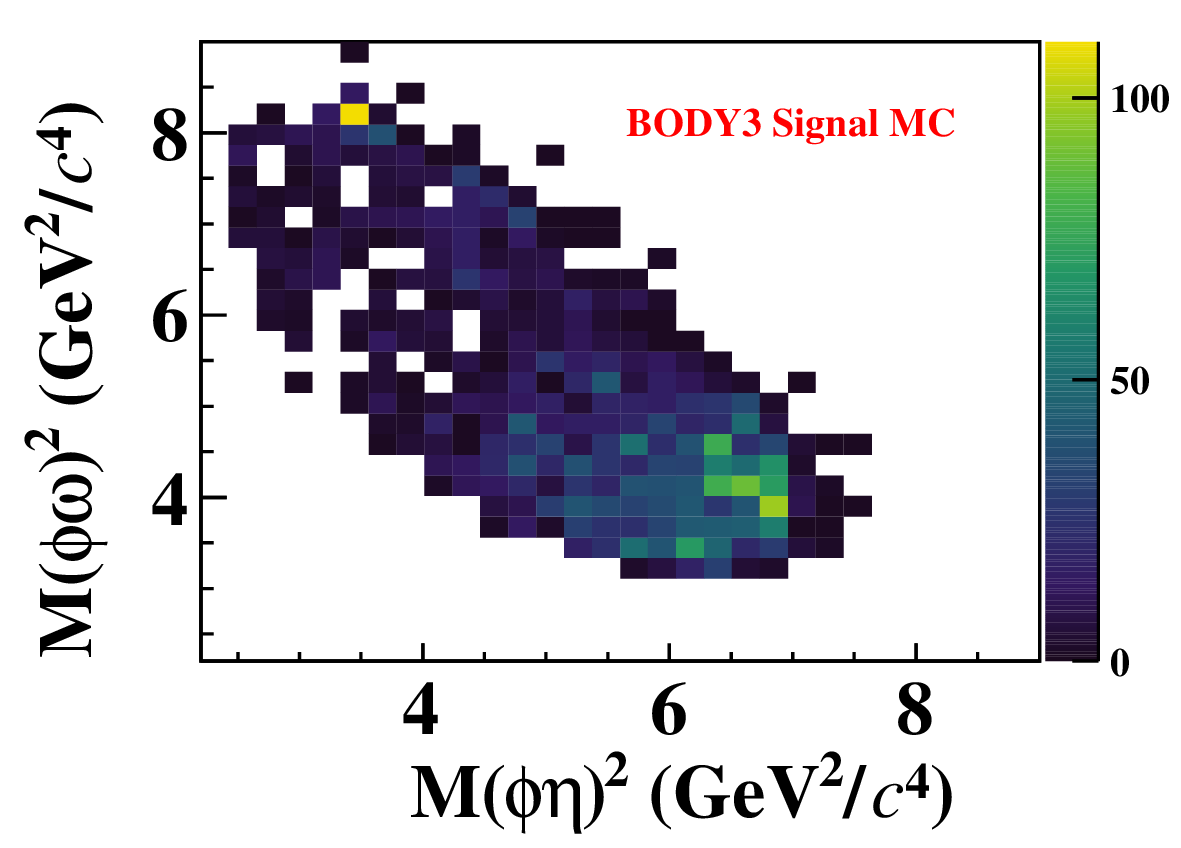}
		\caption{
			Dalitz plots of $M(\phi\omega)^{2}$ versus $M(\phi\eta)^{2}$ in background-subtracted data (first row) and BODY3 MC (second row) for $\chi_{cJ} \to \phi \omega \eta$ candidate events.
			\label{plot:daliz_data_body3}
}
\end{figure*}

The $M(\phi \eta)^{2}$ versus $M(\phi \omega)^{2}$ Dalitz plot distribution in data, corrected bin by bin for background and detection efficiency, is taken as input to the BODY3 generator. 
Figure~\ref{plot:daliz_data_body3} shows the Dalitz plots for the accepted candidates in data and BODY3 MC samples. 
The data-MC comparisons based on the BODY3 signal MC samples are shown in Fig.~\ref{comparision_phi}, where the data-MC agreement is improved compared to that based on the PHSP model.

\subsection{The decay $\chi_{cJ} \to \phi \omega \eta^{\prime}$}
\label{phikkfit}
\begin{figure}[htpt]
\centering
\includegraphics[width=8cm]{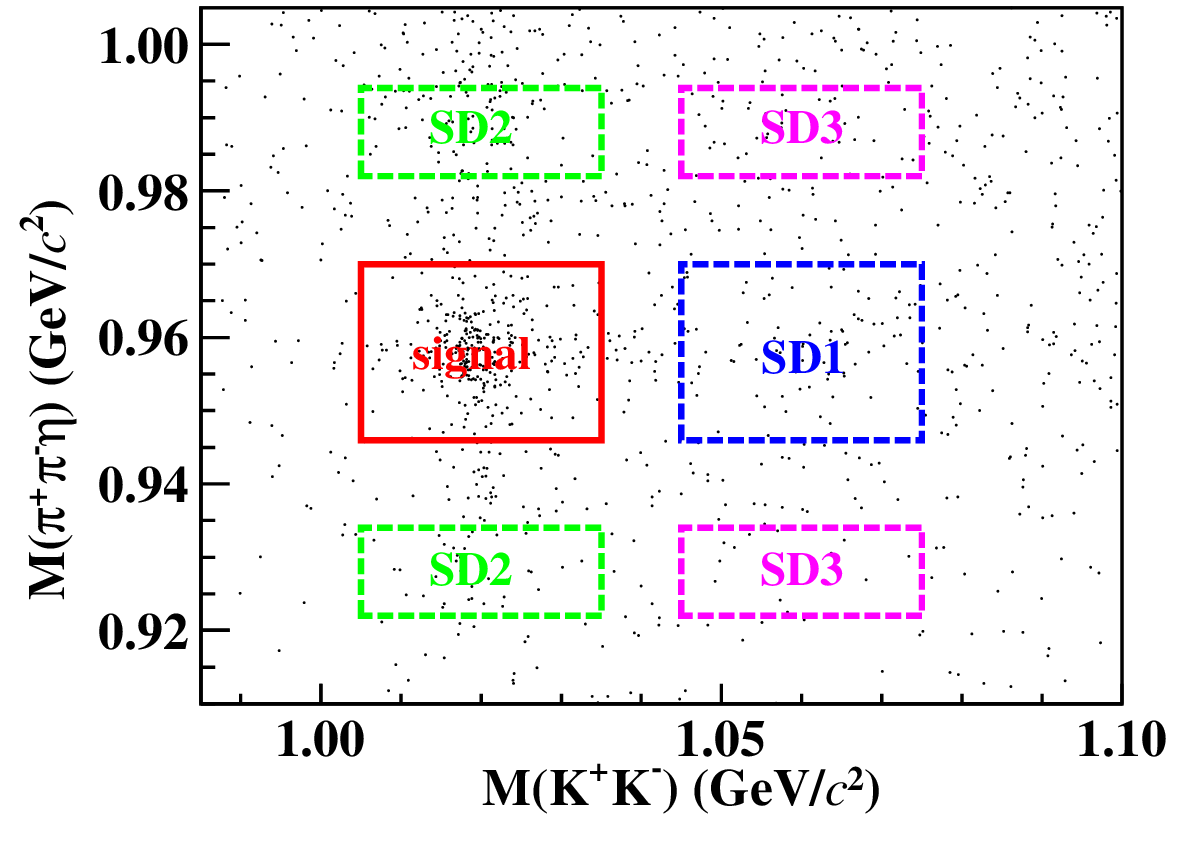}
\caption{Distribution of $M(\pi^{+}\pi^{-}\eta)$ versus $M(K^{+}K^{-})$ for the $\chi_{cJ} \to \phi \omega \eta^{\prime}$ accepted candidates in data, where the red solid box denotes the $\phi \eta^{\prime}$ signal region, the blue dashed box denotes the $\phi \eta^{\prime}$ SD1 region, the green dashed boxes denote the $\phi \eta^{\prime}$ SD2 regions, and the pink dashed boxes denote the $\phi \eta^{\prime}$ SD3 region. Events have been required to be within the $\omega$ signal region, defined as $|M(\pi^{+}\pi^{-}\pi^{0}) - M(\omega)|~ \textless~ 0.024$ GeV/$c^{2}$.
			\label{plot:phietap2D}
		}
\end{figure}

Since in $\chi_{cJ} \to \phi \omega \eta^{\prime}$ events multiple $\pi^{+}\pi^{-}$ combinations are possible, the $\omega$ and $\eta^{\prime}$ are reconstructed by minimizing $\Delta=\sqrt{\frac{\left(M(\pi^{+}_{1}\pi^{-}_{1}\pi^{0})-M(\omega)\right)^2}{{\sigma_\omega}^2}+\frac{\left(M(\pi^{+}_{2}\pi^{-}_{2}\eta)-M(\eta^{\prime})\right)^2}{{\sigma_{\eta^{\prime}}}^2}}$, where $M(\omega)$ and $M(\eta^{\prime})$ are the nominal masses of the $\omega$ and $\eta^{\prime}$ mesons from the PDG~\cite{ref::pdg2025}, and $\sigma_\omega$ and $\sigma_{\eta^{\prime}}$ are the corresponding mass resolutions obtained from signal MC simulation, respectively. The 1D $\eta^{\prime}$ signal region is then defined as $|M(\pi^{+}\pi^{-}\eta) - M(\eta^{\prime})| ~\textless~ 0.012$ GeV/$c^{2}$.
Figure~\ref{plot:phietap2D} shows the 2D distribution of 
$M(\pi^{+}\pi^{-}\eta)$ versus $M(K^{+}K^{-})$
for data after all selection criteria are applied, where the signal region in the red solid box indicates that $\phi$ candidates are required to satisfy $M(K^{+}K^{-}) \in [1.005,\,1.035]\ \text{GeV}/c^{2}$ and $\eta^{\prime}$ candidates are required to satisfy $M(\pi^{+}\pi^{-}\eta) \in [0.946,\,0.970]\ \text{GeV}/c^{2}$. 
Three 2D $\phi \eta^{\prime}$ sideband regions are defined and referred to as ``SD1", ``SD2", ``SD3".
The SD1 region is defined as $M(K^{+}K^{-}) \in [1.045,\,1.075]\ \text{GeV}/c^{2}$ and $M(\pi^{+}\pi^{-}\eta) \in [0.946,\,0.970]\ \text{GeV}/c^{2}$, where $\phi$ candidates lie in the sideband region and $\eta^{\prime}$ candidates in the signal region. The SD2 regions are defined as $M(K^{+}K^{-}) \in [1.005,\,1.035]\ \text{GeV}/c^{2}$ and $0.024 ~\textless~ |M(\pi^{+}\pi^{-}\eta) - M(\eta^{\prime})| ~\textless~ 0.036\ \text{GeV}/c^{2}$, where $\phi$ candidates lie in the signal region and $\eta^{\prime}$ candidates in the sideband region. The SD3 regions are defined as $M(K^{+}K^{-}) \in [1.045,\,1.075]\ \text{GeV}/c^{2}$ and $0.024 ~\textless~ |M(\pi^{+}\pi^{-}\eta) - M(\eta^{\prime})| ~\textless~ 0.036\ \text{GeV}/c^{2}$, where both $\phi$ and $\eta^{\prime}$ candidates lie in the sideband region.

\begin{figure*}[htpt]
	\centering
	\includegraphics[width=14.2cm]{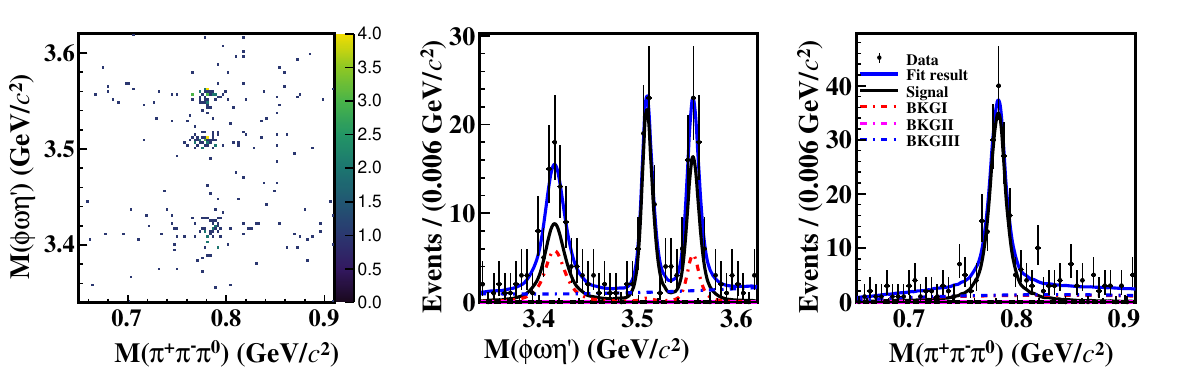}\\
	\includegraphics[width=14.2cm]{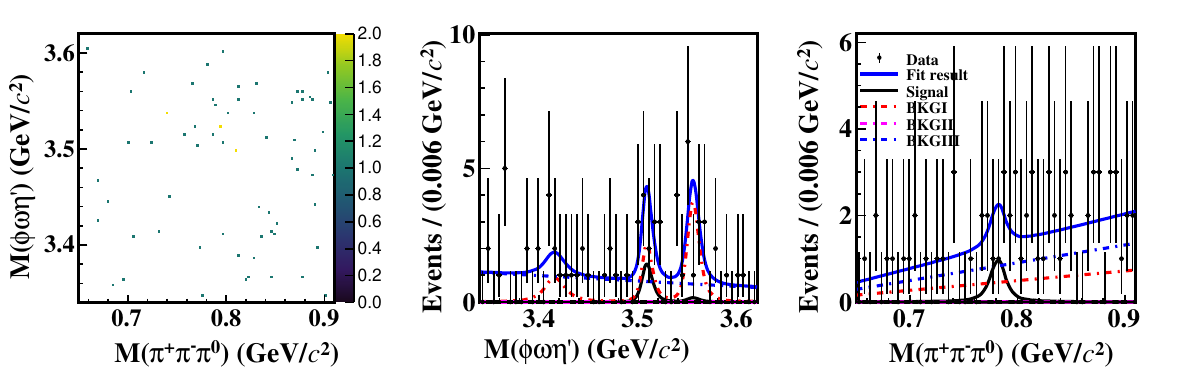}\\
	\includegraphics[width=14.2cm]{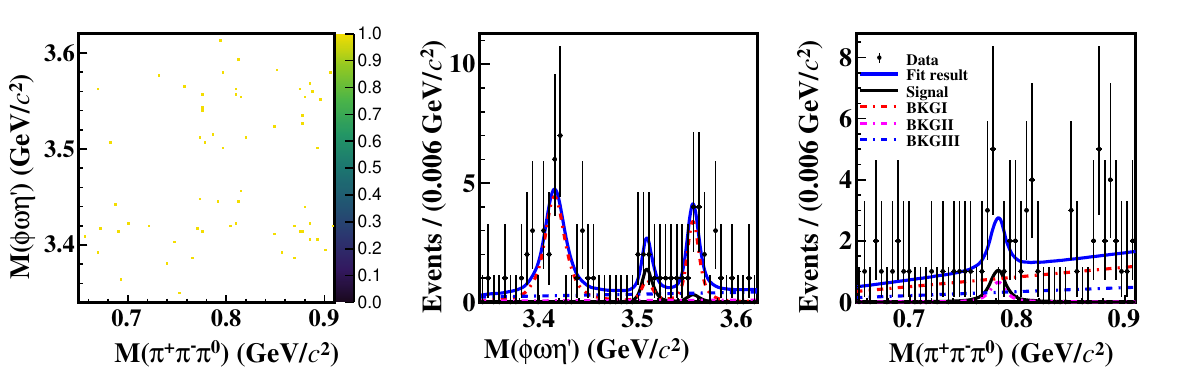}\\
	\includegraphics[width=14.2cm]{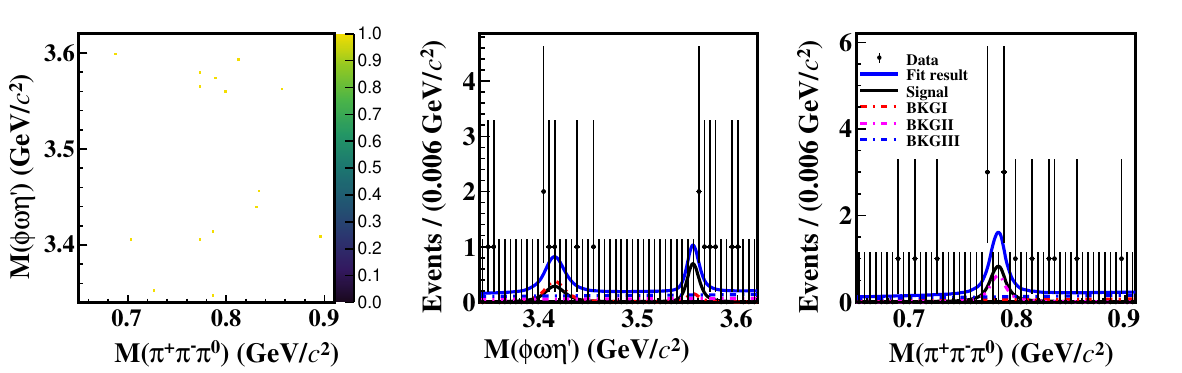}\\
	\caption{Distribution of $M(\phi \omega \eta^{\prime})$ versus $M(\pi^+\pi^-\pi^0)$ (left column), and projections of the 2D fit to the $M(\phi \omega \eta^{\prime})$ (middle column) and $M(\pi^+\pi^-\pi^0)$ (right column) distributions in the $\phi \eta^{\prime}$ signal (first row), sideband-I (second row), sideband-II (third row), and sideband-III (fourth row) regions. For all figures, data are shown as the dots with error bars, the blue solid curves are the fit results, the black solid curves are the fitted signal shapes, the red dashed curves are the BKGI events, the pink dashed curves are the BKGII events, and the blue dashed curves are the BKGIII events.}
	\label{fit:etap}
\end{figure*}

The $\chi_{cJ} \to \phi \omega \eta^{\prime}$ signal yields are determined by a 2D unbinned maximum likelihood fit to the $M(\phi \omega \eta^{\prime})$ versus $M(\pi^{+}\pi^{-}\pi^{0})$ distributions  of accepted candidates in the $\phi$, $\eta^{\prime}$ signal and sideband regions.
In the 2D fit, the PDFs are constructed following the model described in Sec.~\ref{4kfit}. The signal shapes of $\chi_{cJ}$ are from the signal MC-simulated samples, convolved with a Gaussian function; the Gaussian functions in the signal and sideband regions share the width and mean value. The background shape is described by a second-order Chebyshev polynomial .
Figure~\ref{fit:etap} shows the 2D fit results to the signal and sideband candidate events.
No significant non-resonant contribution is observed in the sideband regions. Therefore, the corresponding contribution is neglected in the nominal fit.
From this fit, the $\chi_{c0,1,2} \to \phi \omega \eta^{\prime}$ signal yields are determined, with  statistical significances exceeding $5\sigma$. The obtained signal yields are summarized in Table~\ref{tab:branching_fraction}.

\begin{figure*}[htpt]
\centering
		\includegraphics[width=0.255\textwidth]{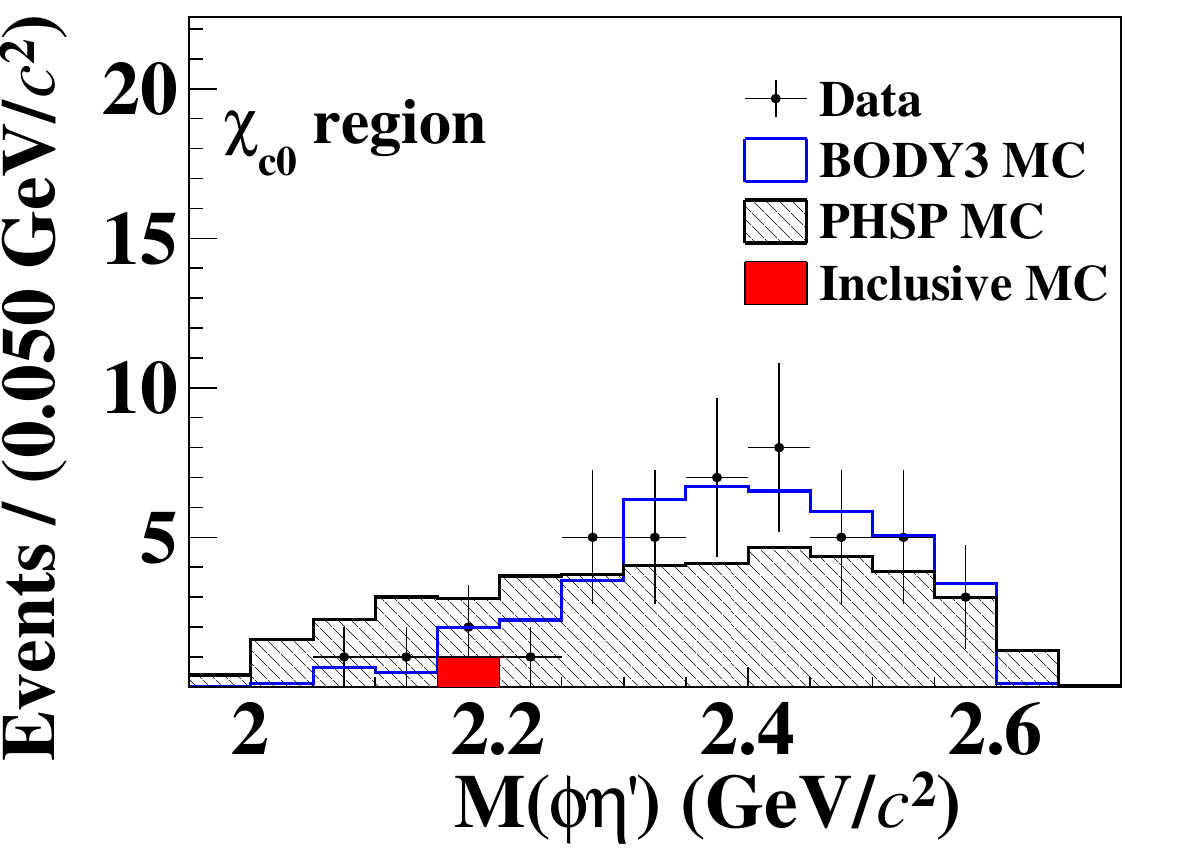}
		\includegraphics[width=0.255\textwidth]{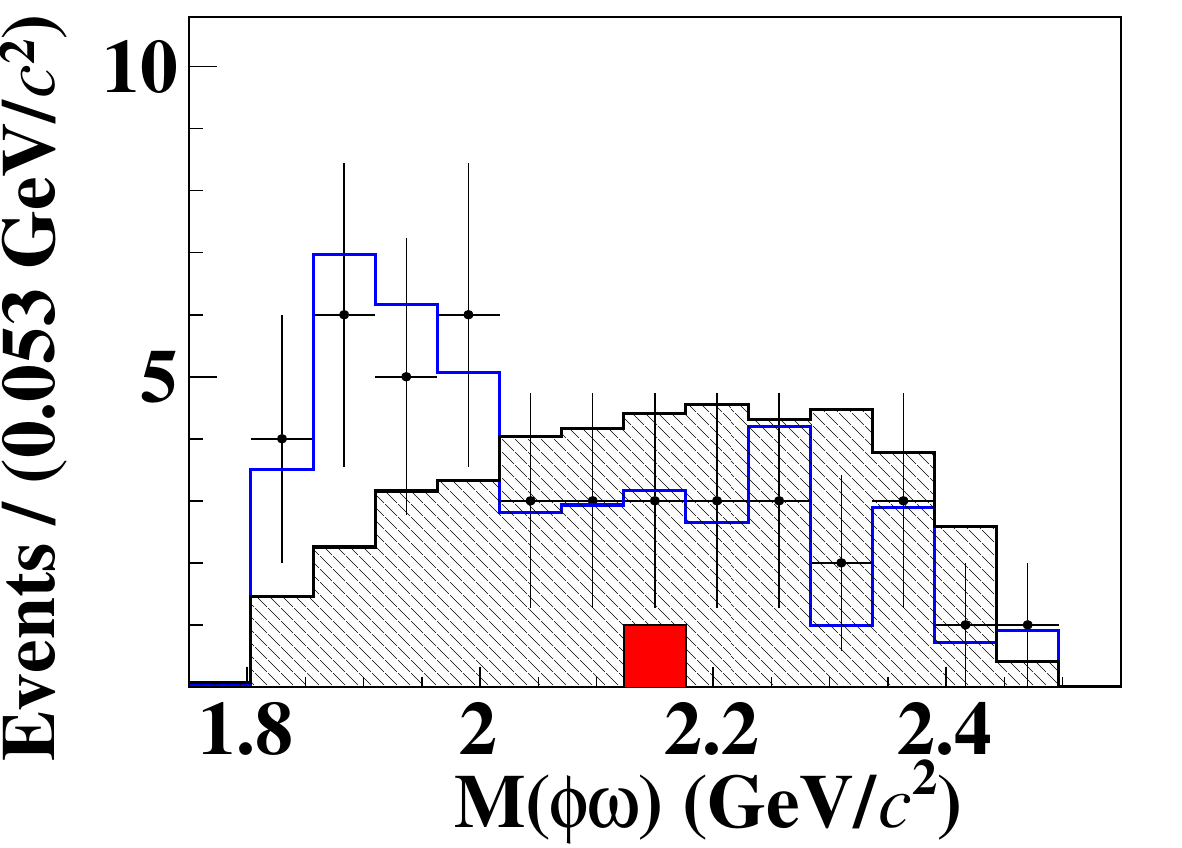}
		\includegraphics[width=0.255\textwidth]{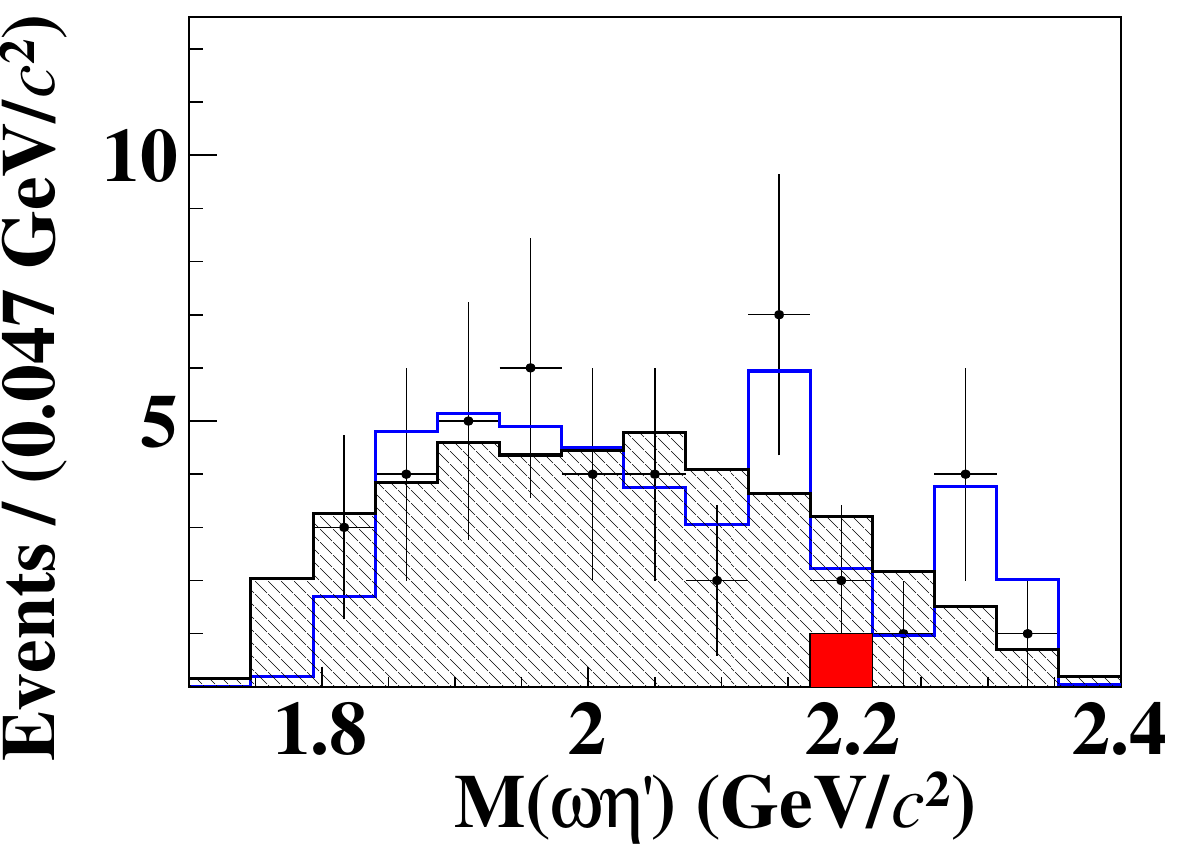}\\
		\includegraphics[width=0.255\textwidth]{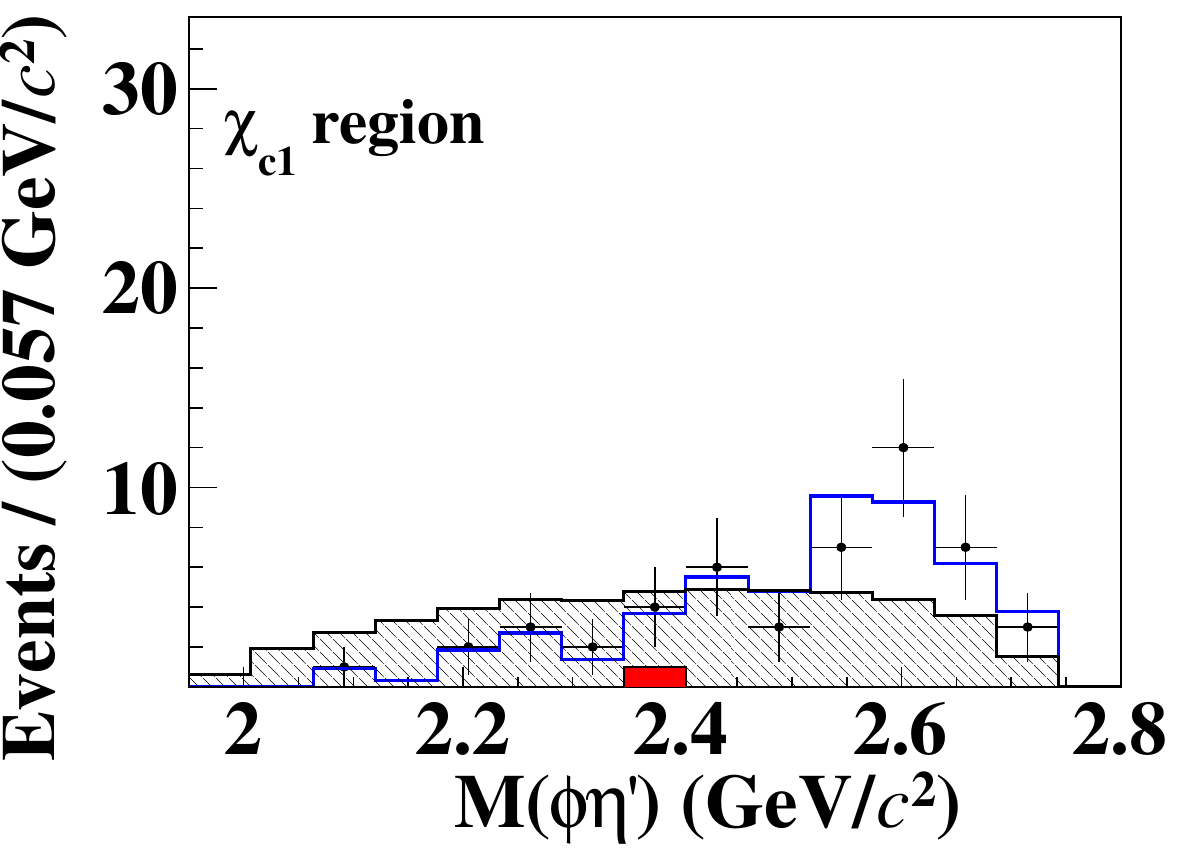}
		\includegraphics[width=0.255\textwidth]{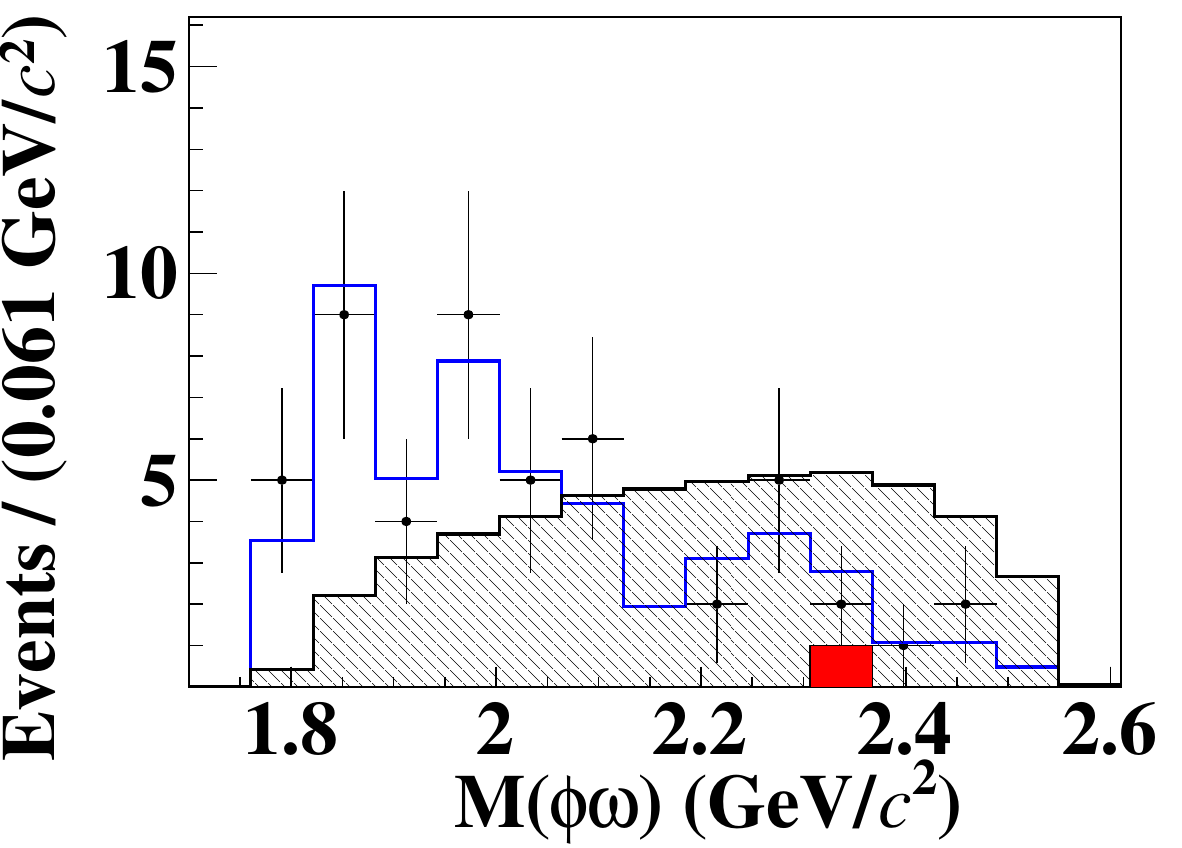}
		\includegraphics[width=0.255\textwidth]{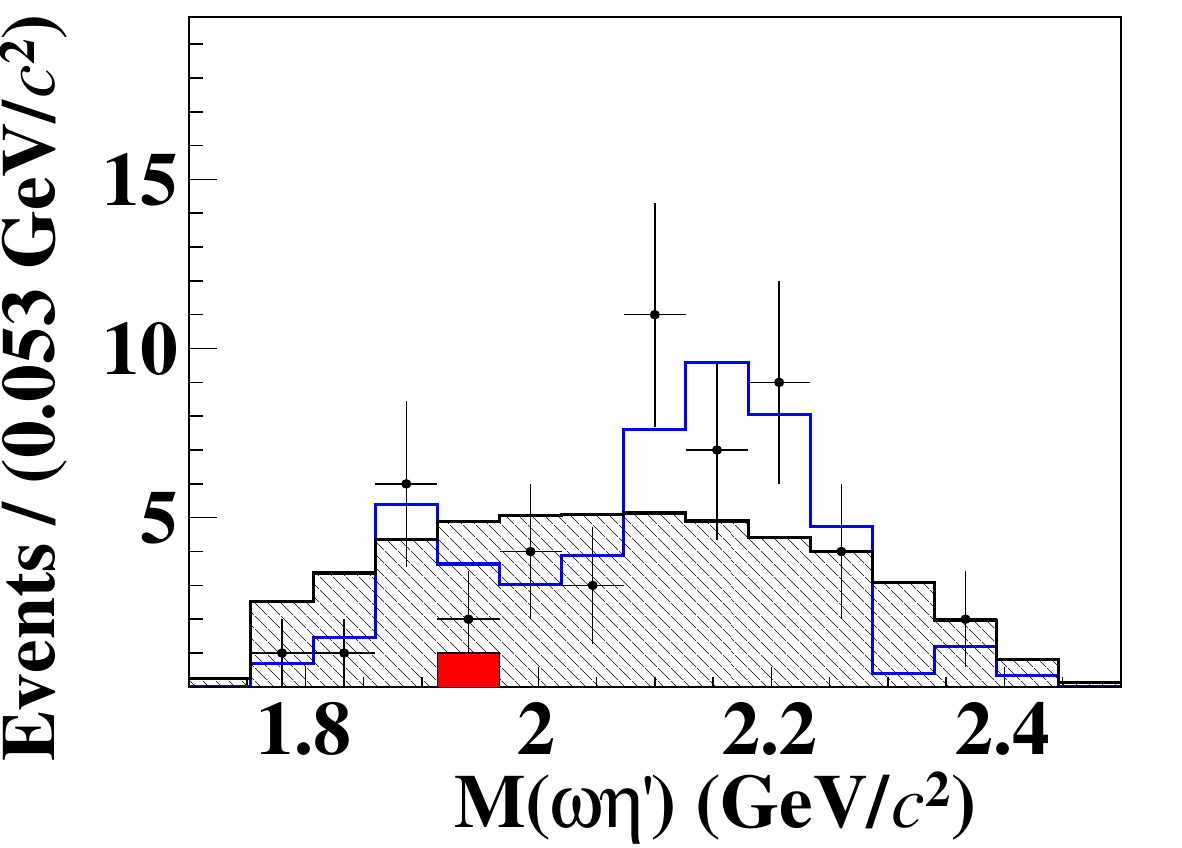}\\
		\includegraphics[width=0.255\textwidth]{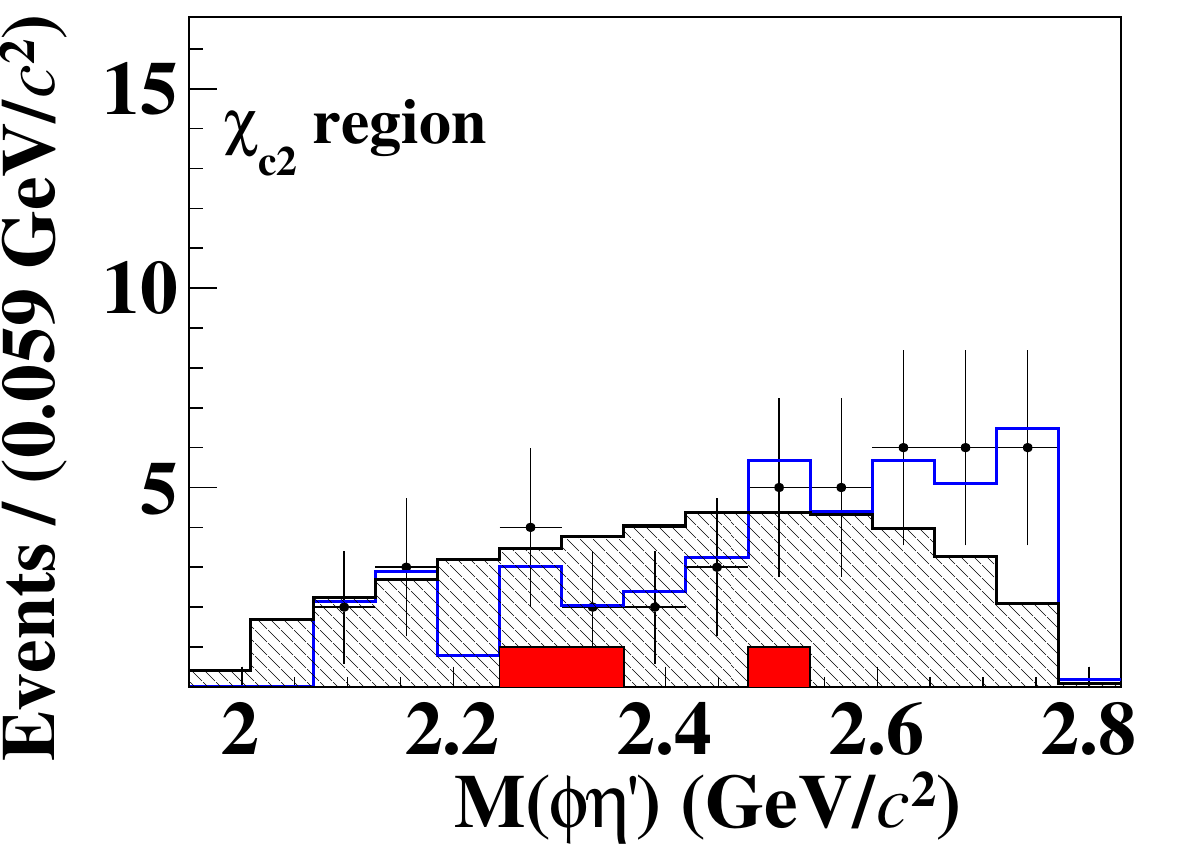}
		\includegraphics[width=0.255\textwidth]{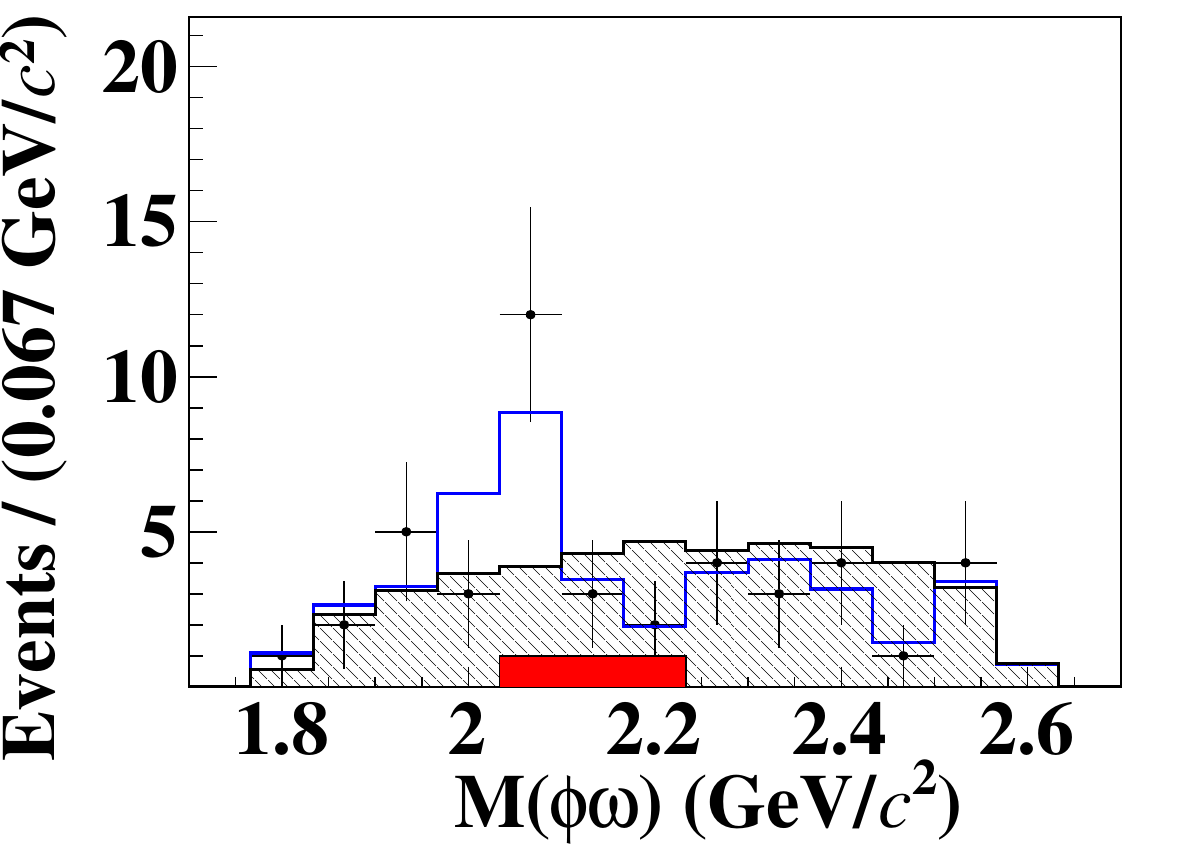}
		\includegraphics[width=0.255\textwidth]{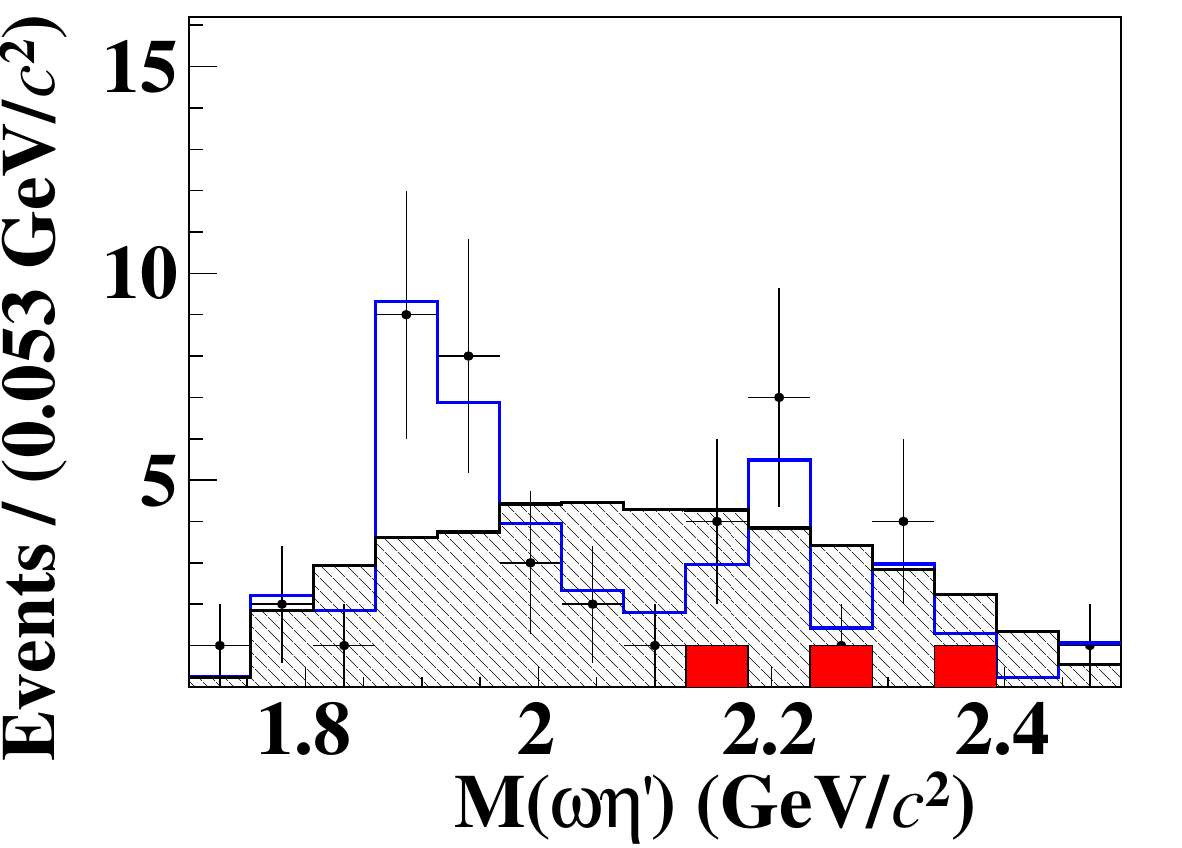}\\
		\caption{
			Invariant mass distributions of possible two-body combinations for the accepted $\chi_{cJ} \to \phi \omega \eta^{\prime}$ candidates. The top, middle, and bottom rows are for events in the $\chi_{c0}$ mass region [3.385, 3.445] GeV/$c^{2}$, the $\chi_{c1}$ mass region [3.490, 3.530] GeV/$c^{2}$, and the $\chi_{c2}$ mass region [3.536, 3.576] GeV/$c^{2}$, respectively. The dots with error bars are data, blue curves are the BODY3 MC samples, red histograms are the background contributions as estimated by the inclusive MC samples, and the black histograms are the PHSP MC samples. The PHSP MC and BODY3 MC samples are normalized to the fitted signal yield in data. Events are required to be within the $\omega$ signal region defined as $|M(\pi^{+}\pi^{-}\pi^{0}) - M(\omega)|~ \textless~ 0.024$ GeV/$c^{2}$.
			\label{plot:comparision_body3_etap}
}
\end{figure*}

\begin{figure*}[htpt]
\centering
\includegraphics[width=0.28\textwidth]{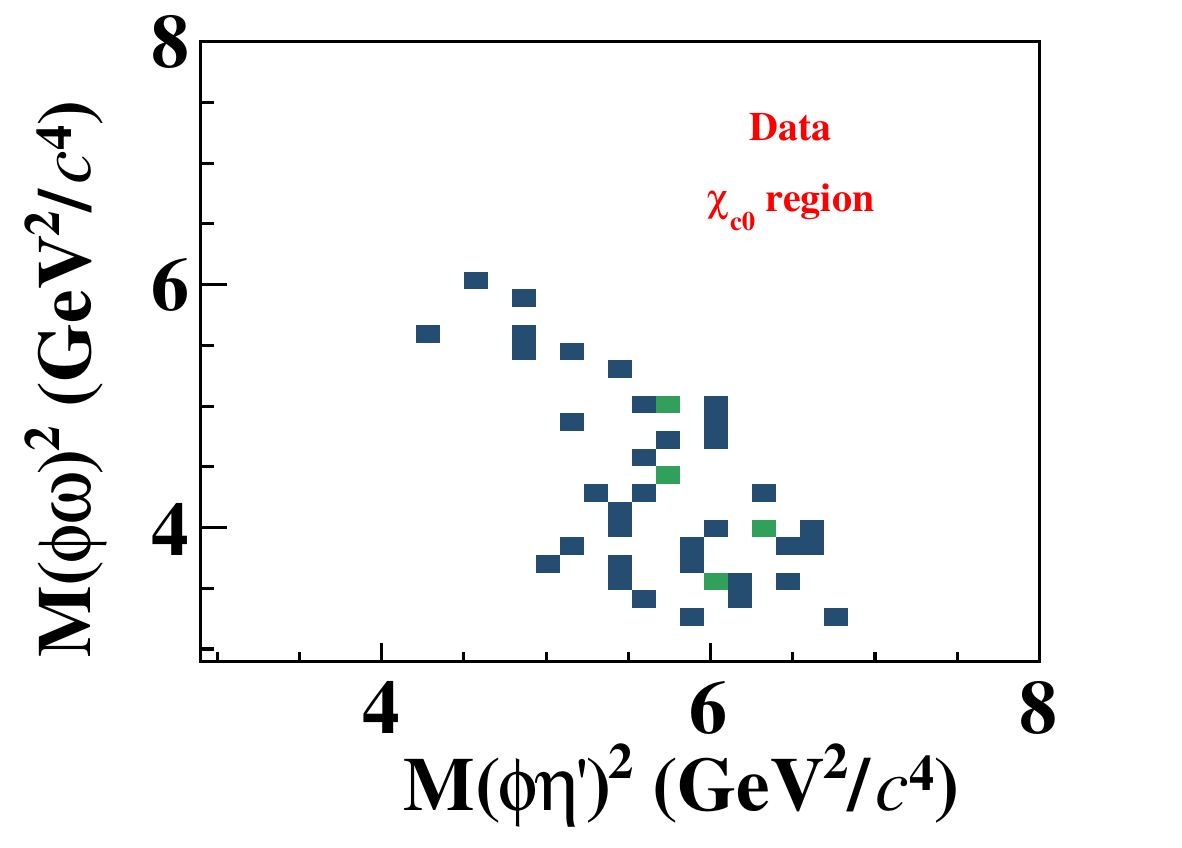}
\includegraphics[width=0.28\textwidth]{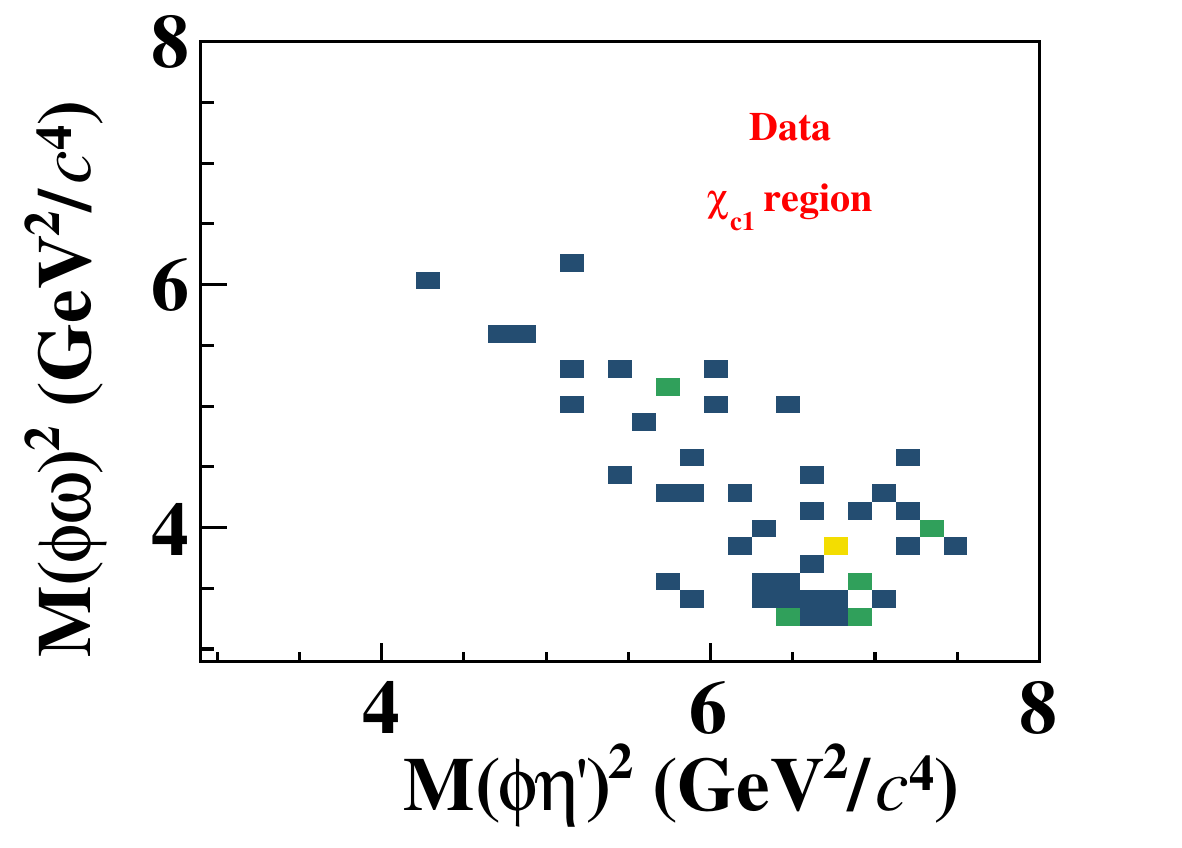}
\includegraphics[width=0.28\textwidth]{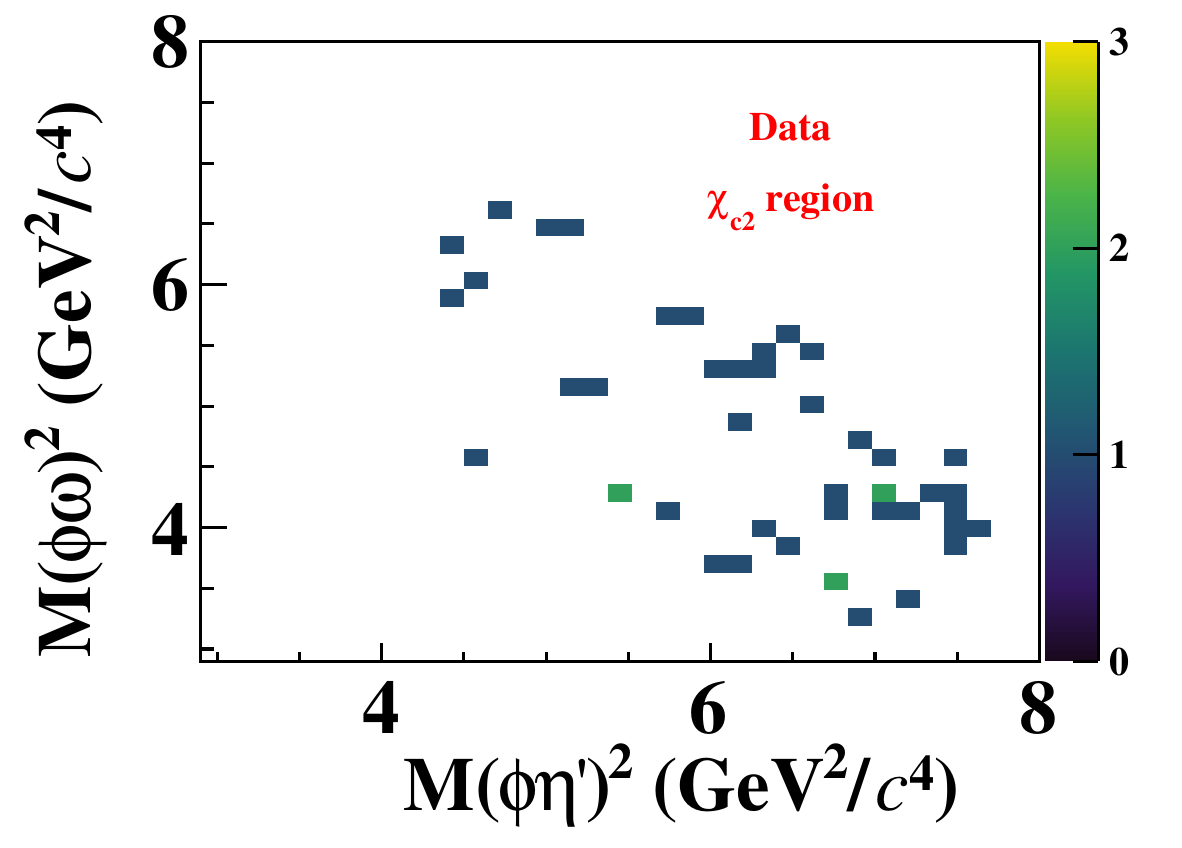}\\
\includegraphics[width=0.28\textwidth]{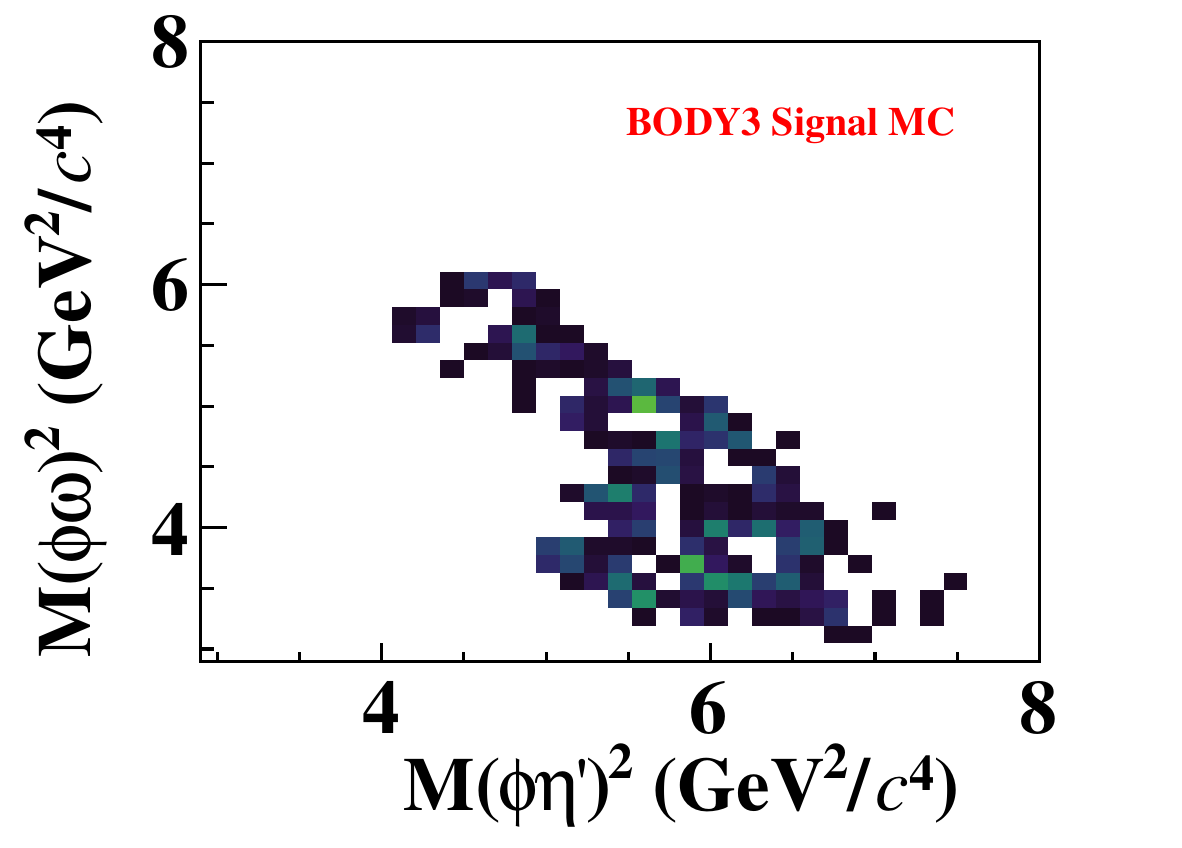}
\includegraphics[width=0.28\textwidth]{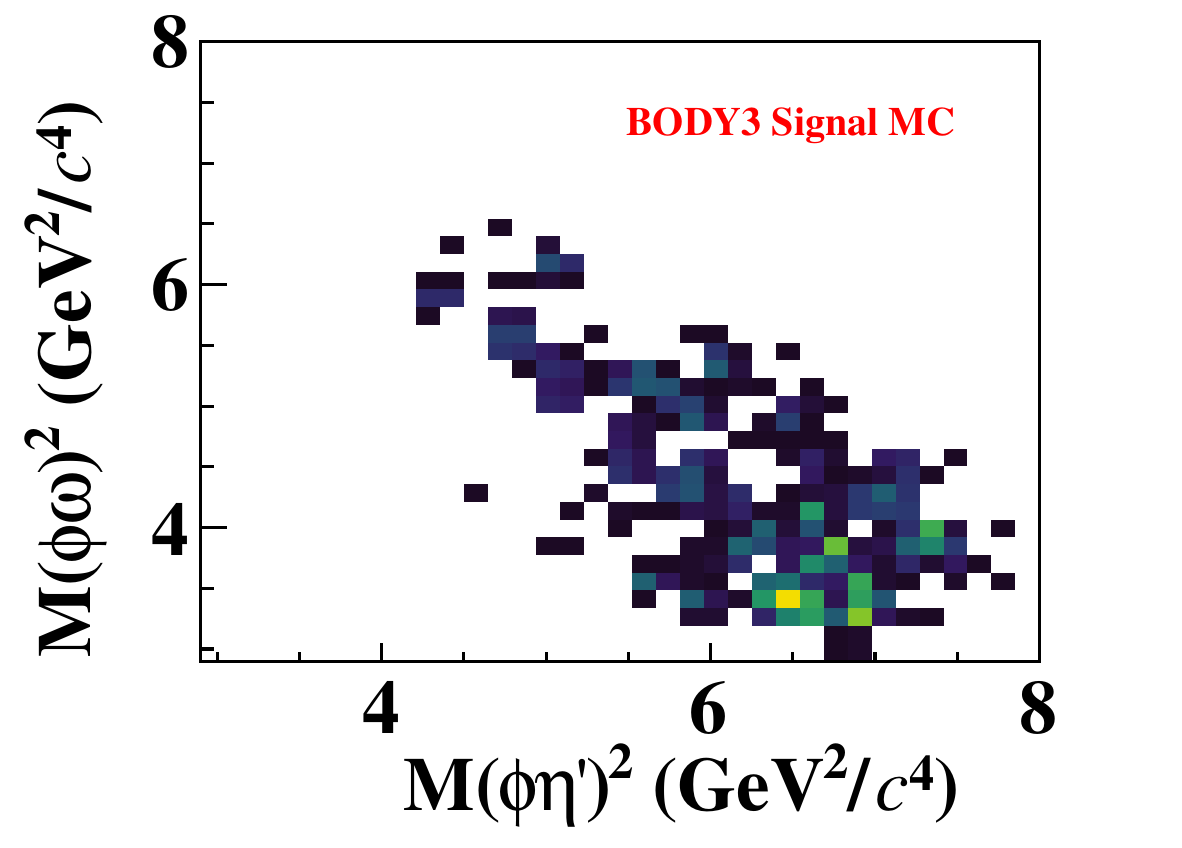}
\includegraphics[width=0.28\textwidth]{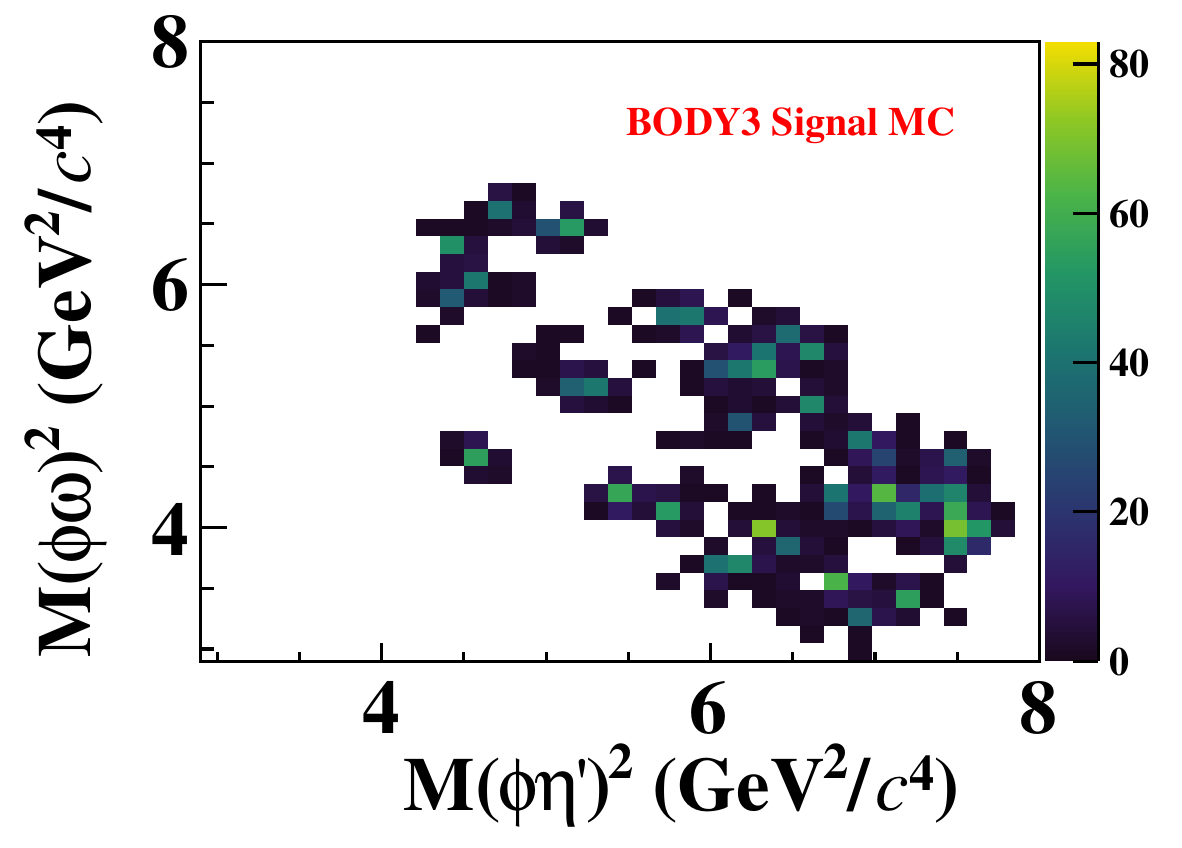}
		\caption{
			Dalitz plots of $M(\phi\omega)^{2}$ versus $M(\phi\eta^{\prime})^{2}$ in data (first row) and BODY3 MC (second row) for $\chi_{cJ} \to \phi \omega \eta^{\prime}$ candidate events.
			\label{plot:daliz_data_body3_etap}
}
\end{figure*}

The existence of potential intermediate states in the $\phi \omega \eta^{\prime}$ final state is investigated through scrutiny of the invariant mass distribution. Figure~\ref{plot:comparision_body3_etap} shows the invariant mass distribution of different two-body combinations for $\chi_{c0,1,2}$ signals, after subtracting the background contributions using the inclusive MC samples. 
Since the observed mass distributions are not well described by the PHSP model, we use the modified data-driven generator BODY3~\cite{Lange01,BODY3} to generate signal MC samples to avoid bias on the efficiency estimation.
The $M(\phi \eta^{\prime})^{2}$ versus $M(\phi \omega)^{2}$ Dalitz plot distribution found in data, corrected bin by bin for background and detection efficiency, is taken as input to the BODY3 generator. 
Figure~\ref{plot:daliz_data_body3_etap} shows the Dalitz plots for the accepted candidates in data and BODY3 MC samples. 
The data-MC comparisons based on the BODY3 signal MC samples are shown in Fig.~\ref{plot:comparision_body3_etap}, where the data-MC agreement is improved compared to that based on the PHSP model.

\subsection{Results}
The BFs of $\chi_{cJ} \to X$ ($X = \phi \omega \eta$, $\phi \omega \eta^{\prime}$) are determined using
\begin{equation}
\begin{aligned}
&\mathcal B(\psi(3686) \to \gamma \chi_{cJ}) \cdot \mathcal B(\chi_{cJ} \to X) =\frac{N_{\rm net}}{N_{\psi(3686)} \cdot \epsilon\cdot{\mathcal B(\rm sub)}  },
\end{aligned}
\end{equation}	
where $\mathcal B(\psi(3686) \to \gamma \chi_{cJ})$ is the BF of $\psi(3686) \to \gamma \chi_{cJ}$, $N_{\psi(3686)}$ is the total number of $\psi(3686)$ events in data and $\epsilon$ is the detection efficiency. $\mathcal B(\rm sub)$ is given by ${\mathcal B}(\phi \to K^+K^-) \cdot {\mathcal B}(\omega \to \pi^{+}\pi^{-}\pi^{0}) \cdot {\mathcal B}(\pi^{0} \to \gamma \gamma) \cdot {\mathcal B}(\eta \to \gamma \gamma)$ for $\chi_{cJ} \to \phi \omega \eta$, and ${\mathcal B}(\phi\to K^+K^-) \cdot {\mathcal B}(\omega \to \pi^{+}\pi^{-}\pi^{0}) \cdot {\mathcal B}(\eta^{\prime} \to \pi^{+}\pi^{-}\eta) \cdot {\mathcal B}(\pi^{0} \to \gamma \gamma) \cdot {\mathcal B}(\eta \to \gamma \gamma)$ for $\chi_{cJ} \to \phi \omega \eta^{\prime}$~\cite{ref::pdg2025}.
Using this formula, the products of the BFs $\mathcal B(\psi(3686) \to \gamma \chi_{cJ}) \cdot \mathcal B(\chi_{cJ} \to X)$ are determined. The $\mathcal B(\chi_{cJ} \to X)$ is obtained by dividing this product by $\mathcal B(\psi(3686) \to \gamma \chi_{cJ})$ taken from the PDG~\cite{ref::pdg2025}.
The obtained results for $\chi_{cJ} \to \phi \omega \eta$ and $\chi_{cJ} \to \phi \omega \eta^{\prime}$ are summarized in Table~\ref{tab:branching_fraction}.

\begin{table*}
    \caption{
     Fitted signal yields, detection efficiencies, and measured BFs for the signal decays. Uncertainties on the yields and efficiencies are statistical only. For the BFs, the first uncertainties are statistical and the second systematic.}
  \centering{
  \begin{tabular}{c|ccc}
    \hline\hline
    & \multicolumn{3}{c}{$\chi_{cJ}\to\phi\omega\eta$} \\
    \hline
    Source
    & $\chi_{c0}$ & $\chi_{c1}$ & $\chi_{c2}$ \\
    \hline
    $N_{\rm net}$
    & $1108\pm46$
    & $631\pm35$
    & $732\pm36$ \\
    $\epsilon$ (\%)
    & $4.30\pm0.03$
    & $5.41\pm0.04$
    & $5.32\pm0.04$ \\
    $\mathcal{B}(\psi(3686)\to\gamma\chi_{cJ})
     \cdot\mathcal{B}(\chi_{cJ}\to\phi\omega\eta)$
     $(\times10^{-5})$
    & $5.47\pm0.23\pm0.50$
    & $2.47\pm0.14\pm0.23$
    & $2.92\pm0.14\pm0.26$ \\
    $\mathcal{B}(\chi_{cJ}\to\phi\omega\eta)$
     $(\times10^{-4})$
    & $5.43\pm0.23\pm0.52$
    & $2.51\pm0.14\pm0.24$
    & $3.03\pm0.15\pm0.28$ \\
    \hline
    & \multicolumn{3}{c}{$\chi_{cJ}\to\phi\omega\eta^{\prime}$} \\
    \hline
    Source
    & $\chi_{c0}$ & $\chi_{c1}$ & $\chi_{c2}$ \\
    \hline
    $N_{\rm net}$
    & $49\pm9$
    & $59\pm9$
    & $51\pm9$ \\
    $\epsilon$ (\%)
    & $1.98\pm0.02$
    & $2.73\pm0.02$
    & $2.70\pm0.02$ \\
    $\mathcal{B}(\psi(3686)\to\gamma\chi_{cJ})
     \cdot\mathcal{B}(\chi_{cJ}\to\phi\omega\eta^{\prime})$
     $(\times10^{-5})$
    & $1.23\pm0.23\pm0.14$
    & $1.08\pm0.16\pm0.13$
    & $0.94\pm0.17\pm0.11$ \\
    $\mathcal{B}(\chi_{cJ}\to\phi\omega\eta^{\prime})$
     $(\times10^{-4})$
    & $1.23\pm0.23\pm0.15$
    & $1.09\pm0.17\pm0.13$
    & $0.98\pm0.17\pm0.11$ \\
    \hline\hline
  \end{tabular}
    \label{tab:branching_fraction}
    }
  \end{table*}

\section{SYSTEMATIC UNCERTAINTY}
\label{sec:systematics}
The systematic uncertainties in the BF measurements originate from several sources, as summarized in Table~\ref{tab:systematics}. They are estimated and discussed below.

\begin{table}[htbp]
	\centering
  	\caption{ Systematic uncertainties in the measurements of the BFs of $\chi_{cJ}\to \phi \omega \eta$ and $\chi_{cJ}\to \phi \omega \eta^{\prime}$ (in \%).}
  \scalebox{0.85}{
    \begin{tabular}{c|c|c} \hline
      \hline         & $\chi_{cJ} \to \phi \omega \eta$ & $\chi_{cJ} \to \phi \omega \eta^{\prime}$    \\
      \hline Source    &$\chi_{c0}~/~\chi_{c1}~/~\chi_{c2}$   &$\chi_{c0}~/~\chi_{c1}~/~\chi_{c2}$\\
      \hline $N_{\psi(3686)}$        &0.5         &0.5              \\
       $(K/\pi)^{\pm}$ PID           &4.0         &6.0              \\
       $(K/\pi)^{\pm}$~tracking      &4.0         &6.0              \\
       $\gamma$~selection            &5.0         &5.0              \\
       $\pi^{0}$ reconstruction      &1.0         &1.0              \\
       $\eta$ reconstruction         &1.0         &1.0              \\
       $\phi$ mass window            &0.1         &0.4                  \\
       $\eta^{\prime}$ mass window   &···         &0.2                  \\
       $J/\psi$ related background   &$2.4~/~2.1~/~0.9$        &···                             \\
       BODY3 generator               &$1.6~/~2.0~/~3.2$        &$4.9~/~1.7~/~3.3$              \\
       Signal yield extraction       &$0.7~/~3.6~/~1.5$        &$2.0~/~5.1~/~2.0$              \\
       Kinematic~fit                 &$3.4~/~2.1~/~2.5$        &$2.1~/~2.2~/~2.5$              \\
       MC~statistics                 &$0.7~/~0.6~/~0.6$        &$1.0~/~0.8~/~0.9$              \\
       ${\mathcal B}(\psi(3686)\to \gamma\chi_{cJ})$   &2.7~/~2.6~/~2.4      &2.7~/~2.6~/~2.4                     \\
       ${\mathcal B}(\rm sub)$       &1.5          &1.9             \\ \hline
      {\bf Total} (without $\psi(3686) \to \gamma \chi_{cJ}$)     &$9.1~/~9.4~/~9.0$         &$11.7~/~11.7~/~11.2$              \\
      {\bf Total}      &$9.5~/~9.7~/~9.3$         &$12.0~/~12.0~/~11.4$              \\
      \hline \hline
    \end{tabular}
}
  \label{tab:systematics}
\end{table}

The total number of $\psi(3686)$ events in data is measured to be $N_{\psi(3686)}=(2712.4 \pm 14.3)\times10^6$ with inclusive
hadronic events, as described in Ref.~\cite{ref::work3}. The uncertainty of $N_{\rm \psi(3686)}$ is 0.5\%.

The systematic uncertainties associated with PID and tracking for $(K/\pi)^{\pm}$ have been studied with the control samples $J/\psi \to K^{*}\bar{K}$ and $J/\psi \to p\bar{p} \pi^{+} \pi^{-}$ in Ref.~\cite{ref::Kpi}. They are assigned as 1.0\% per track for both PID and tracking.

The systematic uncertainty in the photon detection is estimated as 1.0\% per photon with the control sample of $J/\psi\to\pi^+\pi^-\pi^0$. The systematic uncertainty from the $\pi^{0}$ or $\eta$ reconstruction is estimated to be 1.0\% per $\pi^{0}$ or $\eta$ based on the control samples $J/\psi\to\pi^+\pi^-\pi^0$ and $J/\psi \to p \bar{p}\eta$~\cite{ref::gamma-recon}.

The systematic uncertainty associated with the mass window requirements is estimated by studying the differences in mass resolution between the data and signal MC simulation. The differences of the accepted efficiencies between the data and MC simulation are taken as the systematic uncertainties, which are 0.1\% and 0.4\% from the $\phi$ mass window for $\chi_{cJ}\to \phi \omega \eta$ and $\phi \omega \eta^{\prime}$, and 0.2\% from the $\eta^{\prime}$ mass window for the $\chi_{cJ}\to \phi \omega \eta^{\prime}$, respectively.

The systematic uncertainties from the $J/\psi$ veto are estimated by adjusting the relevant mass window. The changes of the net number of events are taken as the corresponding systematic uncertainties, which are 2.4\%, 2.1\% and 0.9\% for $\chi_{c0,1,2} \to \phi \omega \eta$, respectively. No significant background contribution from $J/\psi$ is observed for $\chi_{c0,1,2} \to \phi \omega \eta^{\prime}$ decays. Therefore, the corresponding systematic uncertainty due to the $J/\psi$ background is neglected.

The systematic uncertainties associated with the BODY3 MC model come from the range and the bin divisions of the input Dalitz plot. The uncertainties are estimated by varying the bin size and taking the alternative ranges at the same time. The largest differences in detection efficiencies are assigned as uncertainties, which are 1.6\%, 2.0\%, 3.2\% for $\chi_{c0,1,2}\to \phi \omega \eta$, and 4.9\%, 1.7\%, 3.3\% for $\chi_{c0,1,2}\to \phi \omega \eta^{\prime}$, respectively.

The systematic uncertainties of the signal yields extraction are estimated by examining the fits to the invariant mass spectra. 
The uncertainties from the signal shape are estimated by using alternative Breit-Wigner (BW) functions convolved with a Gaussian function to describe the signal shape, where the Gaussian functions are independent for all four BW functions, and the mass and width of each BW function are fixed to individual PDG values~\cite{ref::pdg2025}. The uncertainties due to the background shape are estimated by varying the orders of the Chebyshev polynomial.
The largest differences between the fit signal yields and the nominal ones are taken as the uncertainties. The uncertainties associated to the invariant mass fit are 0.7\%, 3.6\%, 1.5\% for $\chi_{c0, 1, 2} \to \phi \omega \eta$, and 2.0\%, 5.1\%, 2.0\% for $\chi_{c0, 1, 2} \to \phi \omega \eta^{\prime}$, respectively.

The systematic uncertainty associated with the kinematic fit is estimated by comparing the signal efficiencies before and after applying corrections to the track parameters, with correction factors taken from our previous study~\cite{ref::helixp}.
The changes in signal efficiencies are assigned as the systematic uncertainties,
ranging from 2.1\% to 3.4\% for $\chi_{c0,1,2}\to \phi \omega \eta$ and 2.1\% to 2.5\% for $\chi_{c0,1,2}\to \phi \omega \eta^{\prime}$, respectively, which are summarized in Table~\ref{tab:systematics}.

The systematic uncertainties due to the limited statistics of the signal MC samples are assigned by $\sqrt{\frac{(1-\epsilon)}{N\epsilon}}$, where $\epsilon$ is the detection efficiency and $N$ is the total number of signal MC events.
The resulting uncertainties are 0.7\%, 0.6\%, and 0.6\% for $\chi_{c0,1,2} \to \phi \omega \eta$, and 1.0\%, 0.8\%, 0.9\% for $\chi_{c0,1,2}\to \phi \omega \eta^{\prime}$, respectively.

The systematic uncertainties from the BFs quoted from the PDG~\cite{ref::pdg2025} are 2.7\%, 2.6\%, 2.4\% for $\psi(3686)\to \gamma\chi_{c0,1,2}$, 1.5\% for ${\mathcal B}(\phi \to K^+ K^-) \cdot {\mathcal B}(\omega \to \pi^{+}\pi^{-}\pi^{0}) \cdot {\mathcal B}(\eta \to \gamma \gamma) \cdot {\mathcal B}(\pi^{0} \to \gamma \gamma)$ and 1.9\% for ${\mathcal B}(\phi\to K^+K^-) \cdot {\mathcal B}(\omega \to \pi^{+}\pi^{-}\pi^{0}) \cdot {\mathcal B}(\eta^{\prime} \to \pi^{+}\pi^{-}\eta) \cdot {\mathcal B}(\eta \to \gamma \gamma) \cdot {\mathcal B}(\pi^{0} \to \gamma \gamma)$.

For each signal decay, the total systematic uncertainty is obtained by summing all individual contributions in quadrature.

\section{Summary}
By analyzing a data sample consisting of $(2712.4 \pm 14.3)\times10^6$ $\psi(3686)$ events collected with the BESIII detector operating at the BEPCII collider, 
the decays $\chi_{cJ}\to \phi\omega\eta$ and $\chi_{cJ}\to\phi\omega\eta^{\prime}$ $(J=0,1,2)$  
are measured for the first time with statistical significances greater than $5\sigma$. The BFs are determined to be
\begin{gather*}
\mathcal{B}(\chi_{c0}\to\phi\omega\eta)
  = (5.43\pm0.23\pm0.52)\times10^{-4}, \\[4pt]
\mathcal{B}(\chi_{c1}\to\phi\omega\eta)
  = (2.51\pm0.14\pm0.24)\times10^{-4}, \\[4pt]
\mathcal{B}(\chi_{c2}\to\phi\omega\eta)
  = (3.03\pm0.15\pm0.28)\times10^{-4},
\end{gather*}
and
\begin{gather*}
\mathcal{B}(\chi_{c0}\to\phi\omega\eta^{\prime})
  = (1.23\pm0.23\pm0.15)\times10^{-4}, \\[4pt]
\mathcal{B}(\chi_{c1}\to\phi\omega\eta^{\prime})
  = (1.09\pm0.17\pm0.13)\times10^{-4}, \\[4pt]
\mathcal{B}(\chi_{c2}\to\phi\omega\eta^{\prime})
  = (0.98\pm0.17\pm0.11)\times10^{-4}.
\end{gather*}
Our measurements will be helpful for understanding $\chi_{cJ}\to VVP$ decay mechanisms and exploring previously unobserved $\chi_{cJ}$ decays. These branching fractions can help constrain production amplitudes
in future models incorporating SU(3) flavor symmetry and meson final-state interactions~\cite{ref::Jiang2019,ref::Ikeno2019}. Clear structures are observed in the $\omega \eta$ system; however, due to the limited statistics, we do not investigate these structures in detail. A future data sample with larger statistics, combined with theoretical input incorporating potential interference effects, could improve the understanding of the $\omega \eta$ system.

\section{Acknowledgement}
The BESIII Collaboration thanks the staff of BEPCII (https://cstr.cn/31109.02.BEPC) and the IHEP computing center for their strong support. This work is supported in part by National Key R\&D Program of China under Contracts Nos. 2023YFA1606000, 2023YFA1606704, 2025YFA1613900; National Natural Science Foundation of China (NSFC) under Contracts Nos. 11635010, 11935015, 11935016, 11935018, 12025502, 12035009, 12035013, 12061131003, 12192260, 12192261, 12192262, 12192263, 12192264, 12192265, 12221005, 12225509, 12235017, 12342502, 12361141819, 12535005; the Chinese Academy of Sciences (CAS) Large-Scale Scientific Facility Program; the Strategic Priority Research Program of Chinese Academy of Sciences under Contract No. XDA0480600; CAS under Contract No. YSBR-101; 100 Talents Program of CAS; The Institute of Nuclear and Particle Physics (INPAC) and Shanghai Key Laboratory for Particle Physics and Cosmology; ERC under Contract No. 758462; German Research Foundation DFG under Contract No. FOR5327; Istituto Nazionale di Fisica Nucleare, Italy; Knut and Alice Wallenberg Foundation under Contracts Nos. 2021.0174, 2021.0299, 2023.0315; Ministry of Development of Turkey under Contract No. DPT2006K-120470; National Research Foundation of Korea under Contract No. NRF-2022R1A2C1092335; National Science and Technology fund of Mongolia; Polish National Science Centre under Contract No. 2024/53/B/ST2/00975; STFC (United Kingdom); Swedish Research Council under Contract No. 2019.04595; U. S. Department of Energy under Contract No. DE-FG02-05ER41374


\end{document}